\documentclass[usenatbib]{mn2e}

\usepackage{epsfig,graphicx,amsmath}

\usepackage{amsmath}
\usepackage{lscape}
\usepackage{rotating}
\usepackage{hyperref}
\usepackage[labelsep=period,labelfont=bf]{caption}

\def\apj{ApJ}
\def\apjl{ApJL}
\def\mnras{MNRAS}
\def\pasp{PASP}

\def\araa{ARAA}
\def\aap{A\&A}
\def\aj{AJ}
\def\apjs{ApJS}

\def\gs{\mathrel{\raise0.35ex\hbox{$\scriptstyle >$}\kern-0.6em\lower0.40ex\hbox{{$\scriptstyle \sim$}}}}
\def\ls{\mathrel{\raise0.35ex\hbox{$\scriptstyle <$}\kern-0.6em\lower0.40ex\hbox{{$\scriptstyle \sim$}}}}

\def\Wm2{\,\hbox{W}\,\hbox{m}^{-2}}
\def\gsim{\mathrel{\raise0.35ex\hbox{$\scriptstyle >$}\kern-0.6em\lower0.40ex\hbox{{$\scriptstyle \sim$}}}}
\def\lsim{\mathrel{\raise0.35ex\hbox{$\scriptstyle <$}\kern-0.6em\lower0.40ex\hbox{{$\scriptstyle \sim$}}}}
\newcommand{\appropto}{\mathrel{\vcenter{\offinterlineskip\halign{\hfil$##$\cr\propto\cr\noalign{\kern2pt}\sim\cr\noalign{\kern-2pt}}}}}

\begin{document}

\title[Starburst driven outflows at $z$\,=\,1]{The energetics of
  starburst-driven outflows at $z\sim$\,1 from KMOS}

\author[Swinbank et al.]
{\parbox[h]{\textwidth}
{A.\,M.\ Swinbank,$^{\,1,*}$
C.\, M.\ Harrison,$^{\,1,2}$
A.\ L.\ Tiley,$^{\,1}$
H.\, L.\ Johnson,$^{\,1}$
Ian Smail,$^{\,1}$
J.\,P.\ Stott,$^{\,3}$
P.\, N.\, Best,$^{\,6}$
R.\, G.\ Bower,$^{\,1}$
M.\ Bureau,$^{\,4}$
A.\ Bunker,$^{\,4}$
M.\, Cirasuolo,$^{\,2}$
M.\ Jarvis,$^{\,4}$
G.\, E.\ Magdis,$^{\,5}$
R.\, M.\ Sharples,$^{\,1}$
\& D.\, Sobral$^{\,3}$
}
\vspace*{4pt} \\ 
$^1$Institute for Computational Cosmology, Durham University, South Road, Durham, DH1 3LE, UK\\
$^{*}$Email: a.m.swinbank@dur.ac.uk\\
$^{2}$European Southern Observatory, Karl-Schwarzchild-Str. 2, 85748 Garching b. M{\"u}nchen, Germany\\
$^{3}$Department of Physics, Lancaster University, Lancaster, LA1 4YB, U.K.\\
$^{4}$Sub-dept. of Astrophysics, Department of Physics, University of Oxford, Denys Wilkinson Building, Keble Road, Oxford, OX1 3RH, U.K.\\
$^{5}$Cosmic DAWN Centre, Niels Bohr Institute, University of Copenhagen, Juliane Mariesvej 30, 2100, Copenhagen, Denmark \\
$^{6}$ Institute for Astronomy, University of Edinburgh, Royal Observatory, Edinburgh EH9 3HJ\\
\vspace{-1cm}
}
\maketitle

\begin{abstract}
We present an analysis of the gas outflow energetics from KMOS
observations of 529 main-sequence star-forming galaxies at $z\sim$\,1
using broad, underlying H$\alpha$ and forbidden lines of [N{\sc ii}]
and [S{\sc ii}].  Based on the stacked spectra for a sample with
median star-formation rates and stellar masses of
SFR\,=\,7\,M$_\odot$\,/\,yr and
$M_\star$\,=\,(1.0\,$\pm$\,0.1)\,$\times$\,10$^{10}$\,M$_\odot$
respectively, we derive a typical mass outflow rate of $\dot{M}_{\rm
  wind}$\,=\,1--4\,M$_\odot$\,yr$^{-1}$ and a mass loading of
$\dot{M}_{\rm wind}$\,/\,SFR\,=\,0.2--0.4.  By comparing the kinetic
energy in the wind with the energy released by supernovae, we estimate
a coupling efficiency between the star formation and wind energetics
of $\epsilon\sim$\,0.03.  The mass loading of the wind does not show a
strong trend with star-formation rate over the range
$\sim$\,2--20\,M$_\odot$\,yr$^{-1}$, although we identify a trend with
stellar mass such that
$d$M/$dt$/SFR\,$\propto$\,M$_\star^{0.26\pm0.07}$.  Finally, the line
width of the broad H$\alpha$ increases with disk circular velocity
with a sub-linear scaling relation FWHM$_{\rm
  broad}$\,$\propto$\,$v^{0.21\pm0.05}$.  As a result of this
behavior, in the lowest mass galaxies
(M$_\star\lsim10^{10}$\,M$_\odot$), a significant fraction of the
outflowing gas should have sufficient velocity to escape the
gravitational potential of the halo whilst in the highest mass
galaxies (M$_\star\gsim10^{10}$\,M$_\odot$) most of the gas will be
retained, flowing back on to the galaxy disk at later times.
 \end{abstract}


\section{Introduction}
The process by which massive stars release energy and mass back into
the interstellar medium (ISM) is critical to the evolution of
galaxies.  Theoretical models suggest that ``superwinds'' result when
the energy injection rate from supernovae and stellar winds from OB
associations becomes sufficiently high to excavate a cavity at the
center of the starburst \citep{McCray87}.  The collisions of multiple
stellar winds and supernovae convert their kinetic energy into thermal
energy, and, within the cavity, the hot gas reaches a sound speed much
greater than the local escape velocity building a pressure that is
much higher than the surrounding ISM.  At high pressure gradients the
outflow expands as a ``super-bubble'' along the path of strongest
pressure gradient, sweeping up the ambient medium which collapses into
a thin shell.  The fate of the outflow, which may now be traveling at
hundreds of kilometers per second, depends on several factors
\citep[e.g.\ ][]{Heckman90,Creasey13}.  If the energy injection from
stellar winds and supernovae falls, the expanding shell loses
pressure, cools and eventually stops expanding.  If the starburst is
long lived ($>$10$^{8}$\,yr), then the continual injection of energy
results in a shell that continually expands, although dynamical
and\,/\,or thermal processes may cause the shell to break up into
clumps which allows the interior to freely expand and ``blow out'' of
the galaxy (the so-called ``superwind'').  Since the superwind
preferentially expands along the path of steepest pressure gradient,
the outflows are expected to expand out of the plane of the galaxy, as
seen in local starbursts \citep{Lehnert96}.

The energetics associated with this ``feedback'' process are
critically important for galaxy formation models.  In cosmological
simulations which do not include this stellar feedback, gas rapidly
cools, producing high star-formation rates and the resulting stellar
masses are approximately an order of magnitude higher than observed
\citep[e.g.\ ][]{Somerville99,Cole2k,Benson03,Keres09,Bower12}.  Since
the cold gas fraction in local galaxies is $\lsim$\,5\%
\citep{Saintonge11}, reducing the star-formation efficiency does not
solve the problem, especially in low mass galaxies
\citep{WhiteFrenk91}.  Moreover, the identification of metals within
the lowest density regions of the inter-galactic medium from QSO
absorption line studies \citep{Meyer87,Simcoe04} implies that baryons
can not simply be prevented from entering halos, but rather that some
energetic process must expel baryons from galaxies.

The physical origin and launching mechanism of such a process is
usually attributed to energy and momentum input from supernovae
and\,/\,or radiation pressure from massive stars
\citep{Murray05,Murray10,Krumholz13}.  Both cosmological and
high-resolution ``zoom'' simulations have demonstrated that the
injection of energy from supernovae in the form of kinetic outflows
with constant velocity efficiently removes gas from galaxies, solving
the over-cooling problem by regulating star formation, and enriching
the inter-galactic medium \citep{Springel03,Hopkins12c,Creasey13}.

Observations of local dwarf starbursts and luminous infrared galaxies
(LIRGs) also suggest that the properties of the outflows scale with
those of the parent galaxies
\citep{Lehnert96,Rupke02,Martin05,Westmoquette12}: galaxies with
higher stellar masses and star-formation rates tend to drive higher
velocity outflows, in broad agreement with the momentum-driven wind
model \citep{Murray05}.  In this model, radiation from massive stars
is absorbed by dust which then couples to the gas resulting in
galactic outflows in which the mass loading in the wind scales
proportionally with the star-formation rate \citep{Hopkins12c}.

At high redshift, observations suggest that most star-forming galaxies
are surrounded by ``superwinds''
\citep{Pettini02b,Erb06c,Steidel10,Alexander10,Genzel11,Newman12,Martin12}.
Most of the evidence for these outflows comes from the comparison of
nebular emission line velocities to the absorption lines velocities in
the ISM.  Velocity offsets and broad line widths of several hundred
km\,s$^{-1}$ have been measured, suggesting the outflows are
large-scale \citep[e.g.\ ][]{Steidel10,Freeman19,Forster-Schreiber19}.
Evidence for outflows associated with individual star-forming regions
within high-redshift galaxies also exist
\citep{Newman12b,Genzel11,Genzel14}, most likely locating the source
of the outflows seen on large scales.  However, to determine the mass
outflow rate and kinetic energy in the wind and so provide the
empirical constraints required to test theoretical models, the spatial
geometry of the outflow must be constrained.  To date, most of the
high-redshift galaxies where the winds have been resolved are ``clumpy
disks'' with high star-formation rates, but the ubiquity of winds in
the Lyman-break galaxy study of \citet{Steidel10} suggest they are not
restricted to galaxies with clumpy morphologies.

To address some of these questions, in this paper, we exploit KMOS
observations of the H$\alpha$ emission in 743 star-forming galaxies at
$z\sim$\,1 which were taken as part of the KMOS Redshift One
Spectroscopic Survey (KROSS).  The primary goal of the survey is to
measure the spatially resolved dynamics, velocity dispersion and
metallicities of main-sequence galaxies at $z\sim$\,1
\citep{Stott16,Harrison17}.  Here, we measure the energetics of
starburst driven outflows using the luminosity, line width and spatial
extent of the underlying H$\alpha$ broad line(s) from these galaxies
to measure the ubiquity, kinetic energy and coupling efficiency of the
outflows to the radiation pressure and supernovae.

Throughout the paper, we use a $\Lambda$CDM cosmology
\citep{Spergel07} with $\Omega_{\Lambda}$\,=\,0.73, $\Omega_{\rm
  m}$\,=\,0.27, and $H_{\rm 0}$\,=\,72\,km\,s$^{-1}$\,Mpc$^{-1}$ and
adopt a Chabrier stellar initial mass function (IMF)
\citep{Chabrier03}.  In this cosmology, at $z$\,=\,0.90 (the median
redshift of the galaxies used in this paper), 1$''$ corresponds to a
physical scale of 7.8\,kpc.  All quoted magnitudes are on the AB
system.

%
% Table 1
%

\begin{table*}
{\scriptsize
\begin{center}
\caption{Properties of the narrow and broad lines}
\begin{tabular}{lcccccccccc}
\hline
\hline
                      &       &                     &                    &  \multicolumn{2}{|c|}{Narrow Component}                                                &   \multicolumn{3}{|c|}{Broad Component} \\
Stack            & N$_{\rm gal}$   & $<$\,SFR\,$>$        & $<$\,M$_\star$\,$>$ & FWHM       & \underline{[N{\sc ii}]}  & L$_{\rm H\alpha}$           & FWHM             & \underline{[N{\sc ii}]}  & $\Delta$(BIC) & $\sigma$(P$_{\rm f}$) \\
                      &       & (M$_\odot$\,yr$^{-1}$) & (10$^{10}$\,M$_\odot$) & (km\,s$^{-1}$) &  H$\alpha$             & (10$^{41}$\,erg\,s$^{-1}$) & (km\,s$^{-1}$)   & H$\alpha$ &      &                      \\
\hline
All                  &  529  &  6.7\,$\pm$\,0.2    & 1.0\,$\pm$\,0.1   & 119\,$\pm$\,2   & 0.2\,$\pm$\,0.1    & 2.8\,$\pm$\,0.2   & 280\,$\pm$\,3    & 0.3\,$\pm$\,0.1   & 2947  & 32\\
\hline               
SFR-1                &  89   &  2.4\,$\pm$\,0.1    & 0.8\,$\pm$\,0.1   & 104\,$\pm$\,2   & 0.2\,$\pm$\,0.1    & 1.4\,$\pm$\,0.1   & 181\,$\pm$\,3    & 0.1\,$\pm$\,0.1   & 35    & 5\\
SFR-2                &  108  &  4.3\,$\pm$\,0.1    & 0.8\,$\pm$\,0.1   & 122\,$\pm$\,4   & 0.2\,$\pm$\,0.1    & 1.9\,$\pm$\,0.4   & 202\,$\pm$\,5    & 0.3\,$\pm$\,0.1   & 208   & 13\\
SFR-3                &  110  &  6.4\,$\pm$\,0.1    & 1.0\,$\pm$\,0.1   & 116\,$\pm$\,2   & 0.1\,$\pm$\,0.1    & 2.6\,$\pm$\,0.3   & 231\,$\pm$\,4    & 0.3\,$\pm$\,0.1   & 696   & 22\\
SFR-4                &  111  &  9.2\,$\pm$\,0.1    & 1.1\,$\pm$\,0.1   & 130\,$\pm$\,2   & 0.2\,$\pm$\,0.1    & 2.8\,$\pm$\,0.3   & 247\,$\pm$\,5    & 0.4\,$\pm$\,0.1   & 686   & 23\\
SFR-5                &  111  &  16.2\,$\pm$\,0.8   & 1.5\,$\pm$\,0.2   & 135\,$\pm$\,2   & 0.2\,$\pm$\,0.1    & 4.9\,$\pm$\,0.3   & 277\,$\pm$\,4    & 0.3\,$\pm$\,0.1   & 1346  & 28\\
\hline               
Mass-1               &  100  &  4.8\,$\pm$\,0.6    & 0.2\,$\pm$\,0.1   & 120\,$\pm$\,4   & 0.1\,$\pm$\,0.1    & 1.1\,$\pm$\,0.3   & 180\,$\pm$\,8    & 0.0\,$\pm$\,0.1   & 149   & 9\\
Mass-2               &  124  &  6.4\,$\pm$\,0.4    & 0.6\,$\pm$\,0.1   & 107\,$\pm$\,3   & 0.2\,$\pm$\,0.1    & 3.5\,$\pm$\,0.5   & 197\,$\pm$\,4    & 0.2\,$\pm$\,0.1   & 533   & 19\\
Mass-3               &  124  &  7.1\,$\pm$\,0.7    & 1.1\,$\pm$\,0.1   & 119\,$\pm$\,2   & 0.2\,$\pm$\,0.1    & 4.1\,$\pm$\,0.1   & 212\,$\pm$\,3    & 0.3\,$\pm$\,0.1   & 293   & 14\\
Mass-4               &  99   &  7.8\,$\pm$\,0.8    & 2.0\,$\pm$\,0.1   & 133\,$\pm$\,2   & 0.2\,$\pm$\,0.1    & 3.1\,$\pm$\,0.2   & 277\,$\pm$\,5    & 0.5\,$\pm$\,0.1   & 1344  & 24\\
Mass-5               &  82   &  8.0\,$\pm$\,0.7    & 4.0\,$\pm$\,0.1   & 140\,$\pm$\,2   & 0.2\,$\pm$\,0.1    & 2.5\,$\pm$\,0.2   & 323\,$\pm$\,6    & 0.6\,$\pm$\,0.2   & 2096  & 25\\
\hline
SFRD-1[0.0--0.02]    &  95   &  3.1\,$\pm$\,0.2    & 0.8\,$\pm$\,0.1   & 118\,$\pm$\,6   & 0.2\,$\pm$\,0.1    & 1.2\,$\pm$\,0.4   & 203\,$\pm$\,10   & 0.2\,$\pm$\,0.1   & 112   & 8\\
SFRD-2[0.02--0.05]   &  103  &  4.3\,$\pm$\,0.2    & 1.1\,$\pm$\,0.1   & 112\,$\pm$\,2   & 0.2\,$\pm$\,0.1    & 2.5\,$\pm$\,0.1   & 207\,$\pm$\,3    & 0.3\,$\pm$\,0.1   & 196   & 13\\
SFRD-3[0.05--0.09]   &  109  &  5.9\,$\pm$\,0.3    & 0.9\,$\pm$\,0.1   & 121\,$\pm$\,4   & 0.2\,$\pm$\,0.1    & 2.8\,$\pm$\,0.5   & 210\,$\pm$\,5    & 0.3\,$\pm$\,0.1   & 271   & 15\\
SFRD-4[0.09--0.16]   &  111  &  8.0\,$\pm$\,0.2    & 1.0\,$\pm$\,0.1   & 119\,$\pm$\,2   & 0.1\,$\pm$\,0.1    & 2.6\,$\pm$\,0.2   & 246\,$\pm$\,4    & 0.4\,$\pm$\,0.1   & 1067  & 28\\
SFRD-5[0.16--1.02]   &  111  &  14.7\,$\pm$\,0.6   & 1.3\,$\pm$\,0.1   & 135\,$\pm$\,2   & 0.2\,$\pm$\,0.1    & 5.1\,$\pm$\,0.3   & 268\,$\pm$\,4    & 0.3\,$\pm$\,0.1   & 1074  & 25\\
\hline 
$i$-1[0--36]         &  110  &  6.3\,$\pm$\,0.5    & 1.2\,$\pm$\,0.2   & 106\,$\pm$\,3   & 0.2\,$\pm$\,0.1    & 3.3\,$\pm$\,0.4   & 208\,$\pm$\,4    & 0.3\,$\pm$\,0.1   & 736   & 21\\
$i$-2[36--46]        &  106  &  8.5\,$\pm$\,0.8    & 1.2\,$\pm$\,0.1   & 122\,$\pm$\,3   & 0.1\,$\pm$\,0.1    & 3.2\,$\pm$\,0.4   & 232\,$\pm$\,5    & 0.4\,$\pm$\,0.2   & 784   & 21\\
$i$-3[46--54]        &  106  &  6.0\,$\pm$\,0.4    & 0.7\,$\pm$\,0.1   & 124\,$\pm$\,3   & 0.1\,$\pm$\,0.1    & 2.0\,$\pm$\,0.2   & 235\,$\pm$\,6    & 0.3\,$\pm$\,0.1   & 618   & 17\\
$i$-4[54--65]        &  103  &  7.3\,$\pm$\,0.5    & 1.1\,$\pm$\,0.1   & 129\,$\pm$\,3   & 0.1\,$\pm$\,0.1    & 2.6\,$\pm$\,0.3   & 245\,$\pm$\,6    & 0.3\,$\pm$\,0.1   & 693   & 19\\
$i$-5[65--90]        &  104  &  5.8\,$\pm$\,0.4    & 0.9\,$\pm$\,0.1   & 121\,$\pm$\,4   & 0.1\,$\pm$\,0.1    & 3.0\,$\pm$\,0.5   & 222\,$\pm$\,5    & 0.3\,$\pm$\,0.1   & 390   & 14\\
\hline
sSFR-1[0.06--2.4]    &  92   &  3.8\,$\pm$\,0.3    & 3.0\,$\pm$\,0.2   & 132\,$\pm$\,3   & 0.2\,$\pm$\,0.1    & 1.2\,$\pm$\,0.1   & 279\,$\pm$\,7    & 0.5\,$\pm$\,0.2   & 556   & 18\\
sSFR-2[2.4--4.9]     &  106  &  6.1\,$\pm$\,0.3    & 1.8\,$\pm$\,0.1   & 124\,$\pm$\,2   & 0.2\,$\pm$\,0.1    & 2.6\,$\pm$\,0.2   & 259\,$\pm$\,4    & 0.5\,$\pm$\,0.1   & 743   & 22\\
sSFR-3[4.9--8.8]     &  110  &  7.0\,$\pm$\,0.5    & 1.0\,$\pm$\,0.1   & 112\,$\pm$\,2   & 0.2\,$\pm$\,0.1    & 4.1\,$\pm$\,0.0   & 217\,$\pm$\,3    & 0.3\,$\pm$\,0.1   & 519   & 19\\
sSFR-4[8.8--16]      &  111  &  7.6\,$\pm$\,0.6    & 0.6\,$\pm$\,0.1   & 124\,$\pm$\,3   & 0.1\,$\pm$\,0.1    & 3.0\,$\pm$\,0.5   & 206\,$\pm$\,4    & 0.3\,$\pm$\,0.1   & 309   & 15\\
sSFR-5[1.6--210]     &  110  &  9.7\,$\pm$\,0.6    & 0.4\,$\pm$\,0.1   & 122\,$\pm$\,2   & 0.1\,$\pm$\,0.1    & 2.3\,$\pm$\,0.3   & 214\,$\pm$\,6    & 0.1\,$\pm$\,0.1   & 425   & 17\\
\hline
\label{table:stack}
\end{tabular}
\end{center}
\noindent{\footnotesize Notes: [N{\sc ii}]\,/\,H$\alpha$ denotes
  [N{\sc ii}]$\lambda6583$\,/\,H$\alpha$ flux ratio.  $\sigma$(P$_{\rm
    f}$) denotes the $f$-test significance of the improvement of a
  narrow\,+\,broad line fit over a narrow-only fit.  The units of the
  ranges in column 1 for the star formation rate density (SFRD); inclination ($i$)
  and specific star formation rate (sSFR) are M$_\odot$\,/\,kpc$^2$; degrees
  and 10$^{-10}$\,yr$^{-1}$ respectively.}}
\end{table*}

\section{Analysis \& Results}
\label{sec:analysis}

In this work, we examine galaxies observed as part of the KROSS.  The
sample selection, survey design and data reduction are described in
detail in \citet{Stott16} whilst the dynamical properties and final
sample are described in detail in \citet{Harrison17}.  Briefly, our
KMOS sample comprises observations of 743 star-forming galaxies at
$z\sim $\,1 selected from some of the best studies extra-galactic
survey fields (ECDFS, COSMOS, UDS and SSA\,22).  In the following
analysis, we will stack the spectra of the galaxies in our sample to
search for broad, underlying emission lines.  Before stacking, we
remove the large scale velocity gradients from the galaxy before
creating the galaxy-integrated one-dimensional spectra.  We therefore
limit our analysis to those galaxies that are spatially resolved in
H$\alpha$ emission (although we note that if we include the unresolved
galaxies in the following analysis -- albeit without first removing
the internal velocity gradients from their collapsed spectra -- none
of the quantitative conclusions are significantly changed).  It is
clearly fairer to omit these from the sample so that any broad line
properties are not the result of strong beam-smearing effects.  From
our parent sample of 743 galaxies, 552 are spatially resolved in
H$\alpha$ (see \citealt{Harrison17} for details).

\subsection{Removing (candidate) AGNs from the sample}
\label{sec:AGNs}

In the following analysis, we search for (underlying) broad H$\alpha$
(and [N{\sc ii}] and [S{\sc ii}]) emission from star-formation driven
winds - a signature of out-flowing gas.  Since these lines may be
weak, we stack the spectra of the galaxies.  However, any AGN in our
sample could potentially contaminate the (broad) H$\alpha$ and so next
we identify (and remove) any galaxies with AGN signatures.  The
outflow signatures from AGN-dominated galaxies are discussed in
\citet{Harrison16} and Harrison et al.\ (in prep).

First, we remove 23 AGNs (from the 552 galaxies which are spatially
resolved in H$\alpha$) flagged by \citet{Harrison17}, which are
identified from a [N{\sc ii}]\,/\,H$\alpha$ emission-line ratio of
[N{\sc ii}]\,/\,H$\alpha>$\,0.8 and\,/\,or a $>$\,1000\,km\,s$^{-1}$
H$\alpha$ broad line component in the individual galaxy integrated
spectra.  We verify there are no remaining AGN by investigating the
X-ray luminosities.  We cross correlated the remaining sample with the
\emph{Chandra} X-ray catalogs for their respective fields.  Six of the
galaxies in our remaining sample have X-ray counterparts, all in the
CDFS region and detected in the 2\,Ms catalog of \citet{Luo08}.
However, the 2--10\,keV fluxes of these galaxies, f$_{\rm
  X}$\,=\,2--3\,$\times$\,10$^{-17}$\,erg\,s$^{-1}$\,cm$^{-2}$
correspond to luminosities of $L_{\rm
  X}$\,=\,0.5--1.5\,$\times$\,10$^{41}$\,erg\,s$^{-1}$, which is a
factor $\sim$\,10 below the canonical luminosity expected for the
classification of a source as an AGN ($L_{\rm
  X}\sim$\,10$^{42}$\,erg\,s$^{-1}$).  Since the X-ray emission can be
heavily obscured, finally, we search for obscured AGNs using the
mid-infrared emission \citep[e.g.\ ][]{Stern05} using the
5.8\,--3.6\,$\mu$m versus 8.0\,--4.5\,$\mu$m IRAC colours
which can identify galaxies which are strong power-law sources at
these redshifts \citep[e.g.\ ][]{Donley12}.  However, no additional
candidate AGNs are identified.

Omitting any candidate AGNs and unresolved galaxies from the sample
leaves 529 galaxies for our analysis with well resolved dynamics.  The
median redshift for this sample is $z$\,=\,0.85$^{+0.05}_{-0.03}$ and
the median stellar mass is
log(M$_\star$\,/\,M$_\odot$)\,=\,10.0\,$\pm$\,0.4 with a quartile
range of log(M$_\star$\,/\,M$_\odot$)\,=\,9.5--10.6.  

\subsection{Composite Spectra}
\label{sec:composite}

\begin{figure}
  \centerline{\psfig{file=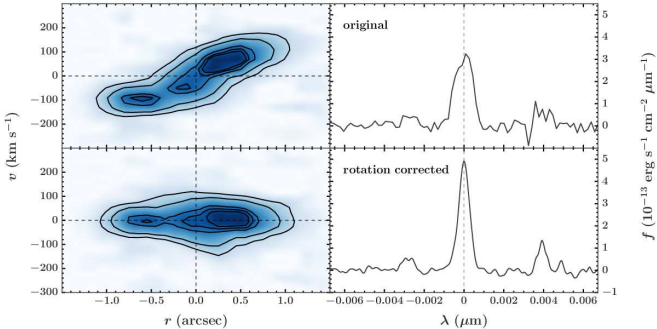,width=3.5in}}
  \caption{Position-velocity diagrams ({\it Left}) and collapsed,
    one-dimensional spectra ({\it Right}) for one of the galaxies in
    our sample.  The top and bottom row shows the position-velocity
    diagram and one-dimensional spectra from the datacube respectively
    before- and after- the large-scale dynamics have been removed.
    The position-velocity diagrams are extracted from the same major
    kinematic axis of the galaxy, and the one-dimensional spectra are
    extracted from the same pixels in both cubes.  This demonstrates
    how removing the large scale galaxy dynamics reduces the line
    width of the galaxy-integrated H$\alpha$ emission.  In this case,
    the width of the H$\alpha$ is reduced from
    $\sigma$\,=\,227\,$\pm$\,11\,km\,s$^{-1}$ to
    $\sigma$\,=\,119\,$\pm$\,9\,km\,s$^{-1}$ in the
    velocity-subtracted cube.}
\label{fig:Vsubtracted}
\end{figure}

Before creating composite spectra, we must remove the large scale
  velocity gradients from the galaxy kinematics which will otherwise
  artificially broaden the observed line emission in the collapsed,
  one-dimensional spectrum of each galaxy (e.g.\ for a disk-like
  system, we need to remove the galaxy's rotation from the cube;
  e.g.\ see \citealt{Harrison17}).  To achieve this, we shift the
  spectra in each pixel to the rest-frame (based on the redshift of
  the H$\alpha$ at that pixel) to create a `velocity-subtracted'
  datacube.  To highlight the effects of this procedure, in
  Fig.~\ref{fig:Vsubtracted} we show the position-velocity diagram and
  collapsed, one-dimensional spectrum for one of the galaxies in our
  sample before and after the galaxy dynamics have been removed from
  the cube.  In this figure, both of the the position-velocity
  diagrams are extracted along the same major kinematic axis of the
  galaxy.  Before the large scale dynamics have been subtracted, the
  rotation of the disk can be seen clearly, with a peak-to-peak
  velocity gradient of 260\,$\pm$\,10\,km\,s$^{-1}$ across $\sim$2$''$
  in projection.  The position-velocity diagram extracted from the
  same kinematic axis of the `velocity-subtracted' cube however shows
  no velocity gradient, demonstrating that the large scale dynamics
  have been removed.  We also show the collapsed, one-dimensional
  spectrum for both cubes, integrated over the same pixels.  This
  demonstrates that removing the large-scale dynamics from the galaxy
  reduces the line width of the H$\alpha$, from
  $\sigma$\,=\,227\,$\pm$\,11\,km\,s$^{-1}$ extracted from the
  original cube to $\sigma$\,=\,119\,$\pm$\,9\,km\,s$^{-1}$ in the
  velocity-subtracted cube.

We caution that it is possible that there may be galaxies in our
sample where the extended H$\alpha$ emission is dominated by an
outflow (e.g.\ a biconical outflow arising from a compact galaxy).  In
this case, removing the velocity gradients will artificially narrow
the broad-line(s) we are searching for.  To test whether this may be
the case, we compare the major kinematic axis (defined from H$\alpha$)
to the major morphological axis (defined from $I$- or $H$-band
\emph{HST} or ground-based $K$-band imaging).  The median offset
between these two position angles is
$\Delta$($|$PA$|$)\,=\,13\,$\pm$\,10$^\circ$ \citep[see
  also][]{Harrison17}).  Thus it appears that the major kinematic axis
and major morphological axis are typically well aligned, which
suggests strong systematic effects from biconical outflows aligned
perpendicular to the disk do not dominate the H$\alpha$ kinematics.

%
%  Figure 1
%

To create a one-dimensional spectrum for each galaxy, we collapse the
velocity-subtracted datacube cube over H$\alpha$ region defined by the
pixel-to-pixel fitting (which reach a median surface brightness limit
of 7\,$\times$\,10$^{-19}$\,erg\,s$^{-1}$\,cm$^{-2}$\,arcsec$^{-2}$).
We then continuum subtract the galaxy spectrum using line free regions
of continuum (defined by regions at least 10\,000\,km\,s$^{-1}$ from
any emission lines) and normalise each galaxy-integrated spectrum by
its H$\alpha$ luminosity to ensure the brightest galaxies do not
dominate the stacks.  A stacked spectrum is created by using a
  sigma-clipped average (using 3$\sigma$ clipping at each wavelength).
  However, we note that all of the results in the following sections
  are insensitive to using a sigma clipped average, inverse
  sky-weighted mean, median or bootstrap average.

\begin{figure*}
  \centerline{\psfig{file=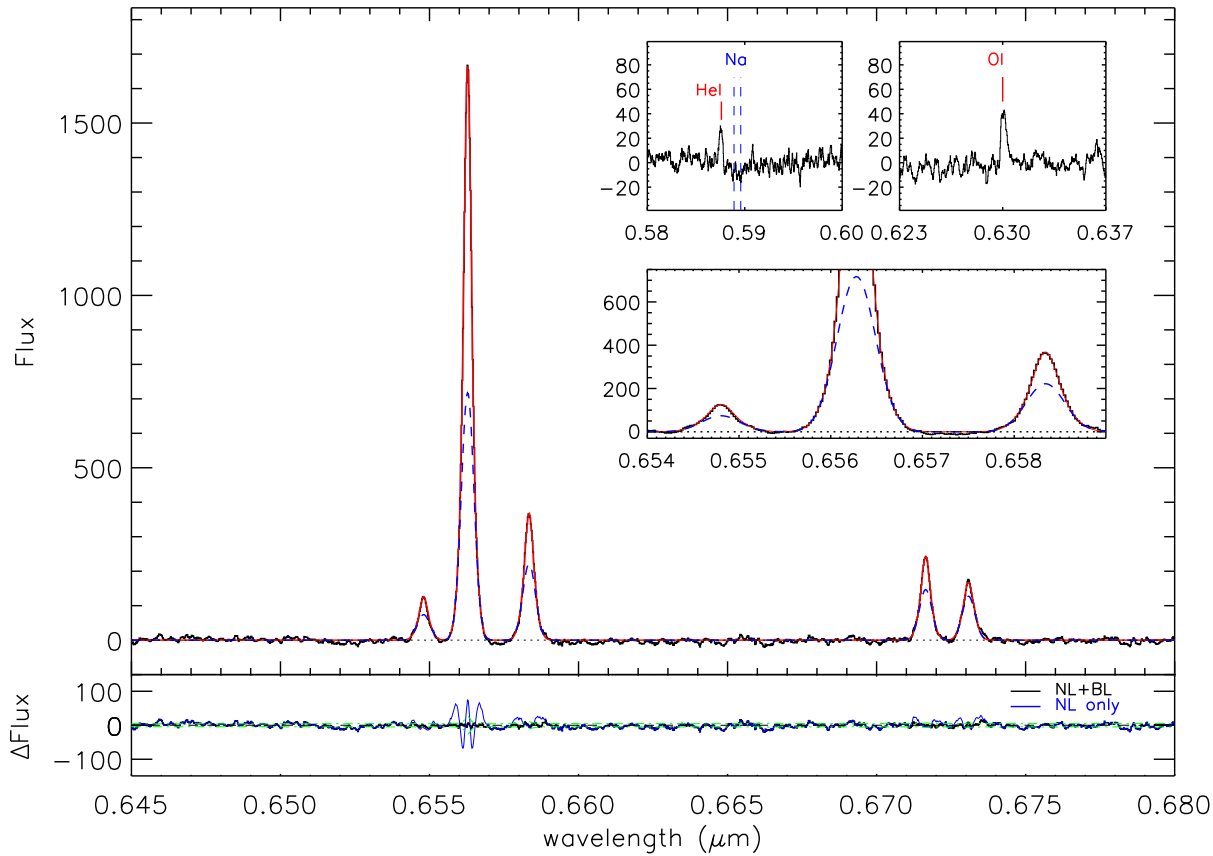,angle=90,width=6in}}
\caption{Rest-frame optical composite of 529 $z\sim$\,1 star-forming
  galaxies from our KMOS observations (black).  We overlay the best
  fit model (red) which comprises both narrow- and broad- emission
  line components (the broad-line component is overlaid as blue
  dashed, and zoomed-up in the central inset).  The upper inset shows
  the detection of the He{\sc i}\,$\lambda$5876 and O{\sc
    i}\,$\lambda$6300 emission, as well as the Na{\sc
    d}$\lambda\lambda$5896,5889 absorption line doublet.  We clearly
  detect strong, broad underlying emission which is evident in the
  H$\alpha$, [N{\sc ii}] and [S{\sc ii}], with
  FWHM\,=\,280\,$\pm$\,3\,km\,s$^{-1}$.  The identification of broad
  emission is suggestive of outflowing material, and since the broad
  emission is seen in the forbidden lines, this suggests that the
  outflow must arise from the ISM (rather than any hidden AGN
  broad-line region).  The lower panel shows the residuals from the
  best-fit (data\,-\,model) for a narrow-emission line fit only (blue)
  and broad+narrow emission line fit (black), which demonstrates that
  the inclusion of the broad component significantly improves the
  fit.}
\label{fig:stack_all}
\end{figure*}

In Fig.~\ref{fig:stack_all} we show the composite spectrum for all of
the galaxies in our sample.  In this stack, we make strong detections
of the H$\alpha$, [N{\sc ii}] and [S{\sc ii}] doublets, He{\sc
  i}\,$\lambda$5876, O{\sc i}\,$\lambda$\,6300 emission lines and weak
Na{\sc D}\,$\lambda$\,5892 absorption.  We note that the effective
integration time in this stack is $\sim$\,3000\,hrs for a galaxy with
an average star formation rate of SFR\,$\sim$\,9\,M$_\odot$\,yr$^{-1}$
and stellar mass of M$_\star\sim$\,10$^{10}$\,M$_\odot$.

%
%  Figure 2
%

\subsection{Searching for Underlying Broad Lines and Gas Outflows}
\label{sec:outflows}

To search for a broad, underlying emission in the composite spectrum,
we use the {\sc mpfit} routine to first fit the narrow-lines of the
H$\alpha$, [N{\sc ii}] and [S{\sc ii}] doublets (allowing the line
widths and normalisations to vary but fixing the [N{\sc ii}] emission
line ratio at [N{\sc ii}]$\lambda$6583\,/\,[N{\sc
    ii}]$\lambda$6543\,=\,2.95; \citealt{Osterbrock82}).  We then
attempt a second fit that includes a broad, underlying emission
component in the [N{\sc ii}], H$\alpha$ and [S{\sc ii}] lines.  
  When fitting a broad component, we couple the broad line FWHM and
  velocity centroid of the underlying [N{\sc ii}], H$\alpha$ and
  [S{\sc ii}].  In Fig.~\ref{fig:stack_all} we show the best
narrow\,+\,broad-line fit to the composite spectra.  We also show the
residual spectra for the narrow-only and narrow+broad line fits.  This
demonstrates that the narrow-line only fit provides a poor fit to the
data compared to the narrow+broad line profile.

To quantify the improved significance of the broad+narrow line fit
compared to the narrow line-only fit, we perform two tests.  First, we
use the BIC statistic; $\Delta$\,BIC\,=\,$(\chi^2_{\rm 1}+k_{\rm 1}
ln(n)) - (\chi^2_{\rm 2}+k_{\rm 2} ln(n))$ where $\chi^2_{\rm 1,2}$
are the total $\chi^2$ of the line fit from the narrow- and
narrow\,$+$\,broad models, $k_{\rm 1,2}$ are the number of degrees of
free parameters in the fit and $n$ is the number of data points.  We
adopt $\Delta$\,BIC$>$10 as strong evidence that a broad line
component is required.  We also perform an $f$-test between the two
fits, with $f$\,=\,($\chi^2_{\rm 1}$\,--\,$\chi^2_{\rm
  2}$)\,/\,($N_{\rm dof,1}$\,--\,$N_{\rm dof,2}$)\,/\,($\chi^2_{\rm
  2}$\,/\,$N_{\rm dof,2}$) and compute the significance of the
improvement using the probability assuming a normal distribution.  We
report both values in Table~1.  In all of the following sections, we
only report the properties of the broad-lines if the
broad\,+\,narrow-line fit provides a significant improvement over the
narrow-only line fit at $>$\,5-$\sigma$ significance.

Since broad, underlying emission could also conceivably be mimicked
from gas motions obeying a Lorentzian distribution, we also attempt a
Lorentzian fit to the stack.  In this fit, we fix the wavelength of
the H$\alpha$ and the doublets of [N{\sc ii}] and [S{\sc ii}], but
allow their width and normalisation to vary (fixing the ratio of
[N{\sc ii}]$\lambda$6583\,/\,[N{\sc ii}]6543\,=\,2.95).  However, the
best-fit Lorentzian profile provides a significantly poorer
($>$\,7-$\sigma$) description of the data in every case.  We therefore
adopt Gaussian profiles as our preferred method of describing the
narrow\,/\,broad components in all of the following sections.

Finally, we note that we have not attempted to correct the
  spectra for Balmer absorption lines in the composite spectra.
  However, to test what effect this might have, we generate synthetic
  stellar spectra using the updated Bruzual-Charlot stellar population
  libraries \citep{Bruzual03}.  We generate a grid of spectra for a
  range of star-formation histories (constant and exponential star
  formation rates with e-folding times of $\tau$\,=\,0.25 0.5, 1, 2,
  5-Gyr) observed at ages that range from 10\,Myr to 8\,Gyr (with 100
  steps that are log-spaced in time) with metallicities ranging from
  0.01 to 2\,Z$_\odot$.  From each spectrum, we measure the equivalent
  width of the H$\alpha$, which has a range of W$_{\rm
    0,H\alpha}$\,=\,$-$1.5 to $-$5.5\,\AA\, with a median of W$_{\rm
    0,H\alpha}$\,=\,$-$4.0\,$\pm$\,0.5\,\AA.  This is much smaller
  than the median equivalent width of the narrow-line H$\alpha$
  emission in an individual galaxy (or stack), which has W$_{\rm
    0,H\alpha}$\,=\,25\AA.  Since the equivalent width of the
  H$\alpha$ emission is stronger than that of the absorption, we have
  not attempted to correct for it here.

\subsection{Properties of the Composite Spectra}
\label{sec:extent}

The inclusion of the broad, underlying component in addition to the
single narrow Gaussian profile in the stacked spectra in
Fig.~\ref{fig:stack_all} provides a significant ($\sim$\,32\,$\sigma$)
improvement over a single Gaussian profile-only fit.  The broad
H$\alpha$ has a FWHM\,=\,280\,$\pm$\,3\,km\,s$^{-1}$ (more than twice
that of the narrow-line H$\alpha$,
FWHM\,=\,119\,$\pm$\,2\,km\,s$^{-1}$).  The narrow-line H$\alpha$ FWHM
is similar to the typical gas-phase velocity dispersion for a
high-redshift disk;
\citep[e.g.\ ][]{Lehnert13,Wisnioski15,Johnson18,Turner17}.
Crucially, the broad emission is not only seen in the H$\alpha$, but
also clearly seen in the forbidden lines of [N{\sc ii}] and [S{\sc
    ii}].  Whilst broad H$\alpha$ is often associated with the
broad-line region of an AGN, the identification of similarly broad
forbidden lines within the spectrum suggest that the broad emission
originates in the lower-density ISM.

To test whether the broad line emission arises only in the nuclear
regions, we create a stack for all pixels which are located
$>$\,0.7$''$ (i.e. outside of 5\,kpc and larger than the seeing disk)
from the galaxy center, and another with all pixels within 0.7$''$ of
the dynamical center.  In both spectra we identify broad, underlying
emission with FWHM\,=\,308\,$\pm$\,6\,km\,s$^{-1}$ and
FWHM\,=\,286\,$\pm$\,7\,km\,s$^{-1}$ for the $r<$\,0.7$''$ and
$r>$\,0.7$''$ apertures respectively.  This suggests that the broad,
underlying emission is not confined to the nuclear regions, but is a
galaxy-wide phenomenon (we discuss this further in
Section~\ref{sec:SpatialExtent}).

As Fig.~\ref{fig:stack_all} shows, the composite spectrum shows strong
[N{\sc ii}], H$\alpha$ and [S{\sc ii}] emission, but we also detect
the He{\sc i}\,$\lambda$5876 and (low ionisation) O{\sc
  i}\,$\lambda$6300 emission lines with ratios of O{\sc
  i}\,/\,H$\alpha$\,=\,0.036\,$\pm$\,0.003 and He{\sc
  i}\,/\,H$\alpha$\,=\,0.020\,$\pm$\,0.002 (in these calculations
  we use the total H$\alpha$ emission line flux since there is
  insufficient signal-to-noise to decompose the He{\sc i} and [O{\sc
      i}] in to narrow and broad components).  The [O{\sc
    i}]\,/\,H$\alpha$ emission line ratio provides a diagnostic of the
average ISM conditions.  The [O{\sc i}]\,/\,H$\alpha$ emission line
ratio is relatively weak in gas photo-ionised by stars ([O{\sc
    i}]\,/\,H$\alpha\lsim $\,0.08) and stronger in shock-heated gas
and AGN \citep{Osterbrock89,Kewley06}.  Indeed, in the classification
system of \citet{Kewley06}, ``transition'' (or composite) galaxies and
AGN are classified by O{\sc i}\,/\,H$\alpha$\,$\gsim$\,0.06 and 0.15
respectively with all galaxies with O{\sc
  i}\,/\,H$\alpha$\,$\lsim$\,0.05 classed as H{\sc ii} region-like
(this is also the case if we use the total [O{\sc i}] flux and narrow
H$\alpha$ flux).  Thus, the line ratios suggest that the ionisation of
the ISM in our sample is dominated by photoionisation of the gas.

An important quantity for measuring the ionised gas mass and hence
mass outflow rates is the electron density ($n_e$).  This quantity is
often measured from the ratio of the [S{\sc ii}]\,$\lambda$6716,6731
doublet.  The forbidden lines of [S{\sc ii}]\,$\lambda$6716,6731 arise
in low-density gas, where there are too few collisions to de-excite
the excited state.  In the low density regime
($\lsim$\,10\,cm$^{-3}$), the ratio of the [S{\sc
    ii}]$\lambda$6716\,/\,[S{\sc ii}]$\lambda$6731 provides a
statistical weight, reflecting how many electrons the level can hold.
At high density ($\gsim$10$^4$\,cm$^{-3}$), the ratio instead reflects
the shorter lifetime of the $\lambda$6731\AA\ transition.  In between,
the ratio measures the gas density \citep[e.g.][]{Osterbrock06}.

From the integrated spectra (Fig.~\ref{fig:stack_all}), we measure a
narrow-line ratio of [S{\sc ii}]$\lambda$6716\,/\,[S{\sc
    ii}]$\lambda$6731\,=\,1.4\,$\pm$\,0.1, which implies an electron
density of $n_{\rm e}$\,=\,30$_{-20}^{+40}$\,cm$^{-3}$, and a broad
line ratio of [S{\sc ii}]\,6716\,/\,[S{\sc
    ii}]$\lambda$6731\,=\,1.3\,$\pm$\,0.1, which suggests an electron
density of $n_{\rm e}$\,=\,75$_{-50}^{+55}$\,cm$^{-3}$
\citep{Osterbrock06}.

\subsection{Spatial Extent and Limits on the Geometry of the Outflows}
\label{sec:SpatialExtent}

\begin{figure}
  \centerline{\psfig{file=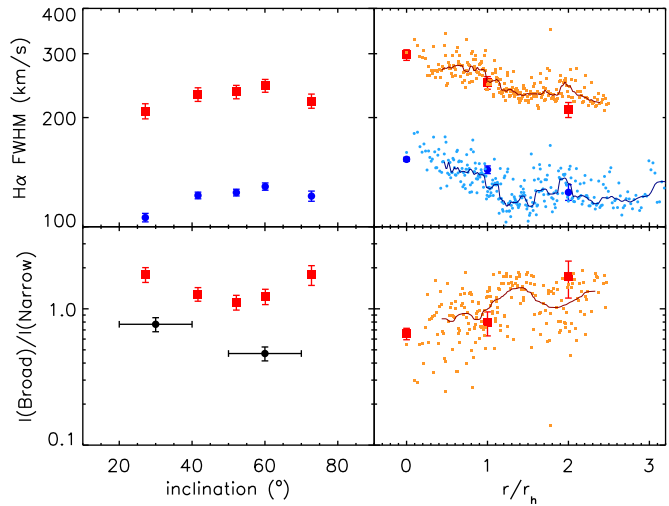,angle=90,width=3.5in}}
  \caption{H$\alpha$ broad- and narrow- line properties as a function
    of inclination and radius.  {\it Top Left:} H$\alpha$ FWHM for the
    narrow (circle) and broad (square) Gaussian profiles as a function
    of galaxy inclination.  The data show a weak trend of increasing
    broad line FWHM with inclination, from 208\,$\pm$\,4\,km\,s$^{-1}$
    for galaxies at low inclination, to 245\,$\pm$\,6\,km\,s$^{-1}$
    for more highly inclined galaxies.  Although the trend is weak, it
    is consistent with several other studies which have also shown
    that outflows tend to be bi-polar with wide opening angles
    \citep{Martin12,Bordoloi11}.  {\it Top Right:} H$\alpha$ FWHM for
    the broad and narrow Gaussian profile as a function of radius
    (normalised to half light radius).  The large symbols show the
    FWHM in three radial bins, whilst the small points show the FWHM
    in bins that each contain 100 spectra from the datacubes, ranked
    by their (normalised) radius (in units of $r$\,/\,r$_{\rm h}$).
    The solid lines denote running medians through the data.  {\it
      Bottom Left:} H$\alpha$ broad\,/\,narrow-line luminosity ratio
    as a function of inclination. The solid circles show the intensity
    ratio of broad-to-narrow line intensities from the SINS survey
    \citep{Newman12}.  {\it Bottom Right:} H$\alpha$
    broad\,/\,narrow-line luminosity ratio as a function of radius
    (normalised to half light radius).  The small points show the FWHM
    in bins that each contain 100 spectra from the cubes, with the
    solid line showing a running median through the data,
    demonstrating that the luminosity ratio of broad- to narrow-line
    H$\alpha$ flux increases with half light radii.}
\label{fig:stack_radius}
\end{figure}

Before we can use these data to estimate properties of the high
velocity gas (such as the potential mass in the wind, the mass loading
and its kinetic energy), we first require an estimate of the typical
spatial extent and geometry of the outflows.  We therefore stack the
galaxy spectra in annulli (adopting the dynamical center from the
best-fit disk model) and search for broad, underlying emission
(Fig.~\ref{fig:stack_radius}).  To account for differences in galaxies
sizes in the sample, in Fig.~\ref{fig:stack_radius} we parameterise
the intensity and intensity ratio and FWHM of the broad H$\alpha$ in
terms of half-light radii of the galaxies.  The spatial extent of the
broad H$\alpha$ emission appears to be extended over at least 2.5 half
light radii (the median half light radius of the sample is
2.7\,$\pm$\,0.1\,kpc with a quartile range of 1.8--3.4\,kpc).  As a
further test, we also bin all the spectra from the cubes ranked by
their (normalised) radius (in top sets of 100 spectra) and then fit
each of the stacks and measure the broad and narrow line intensity
ratios and FWHM.  We also overlay these results in
Fig.~\ref{fig:stack_radius}, which show good agreement with the
results from coarser bins, and suggests that the outflows are not
confined to the central regions, but extend across several half-light
radii.

Since some of the galaxies are small (and so normalising by half light
radii may introduce artificial effects from the seeing), we also
measure the spatial extent of the broad H$\alpha$ in radial bins in
arcseconds (without renormalising), and estimate a spatial FWHM of the
broad-line intensity of 1.5\,$\pm$\,0.2$''$ (10\,kpc at $z$\,=\,1).
This corresponds to $\sim$\,10\,kpc at $z\sim$\,1 further suggesting
that the outflows are not confined to the nuclear regions.  It is
interesting to note that the FWHM of the broad line falls over the
range 1--10\,kpc (Fig.~\ref{fig:stack_radius}).  This is similar to
local starbursts, where the FWHM from the outflowing gas along the
major axis of the galaxy falls rapidly with radius
(e.g.\ \citealt{Lehnert96}; although note that the disks studied in
\citet{Lehnert96} have half light radii which are a factor
$\sim$\,2\,$\times$ larger than those studied here).

%
%  Figure 3
%

\begin{figure}
  \centerline{\psfig{file=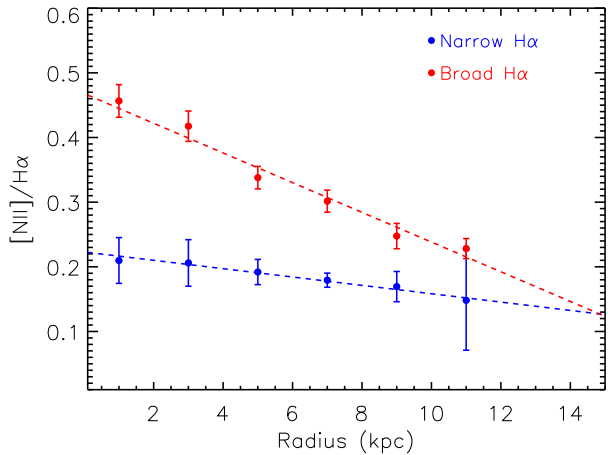,angle=90,width=3.3in}}
  \caption{ [N{\sc ii}]\,/\,H$\alpha$ emission line profiles for the
    narrow- and broad- emission lines in the stacked spectra as a
    function of galacto-centric radius.  Both the narrow- and broad-
    emission line ratios decline with increasing radius, although the
    narrow-line gradient is much weaker (but consistent with that
    typically detected in individual high-redshift galaxies). }
\label{fig:NIIHa_radius}
\end{figure}

In the KMOS sample, we also detect strong variation in the emission
line ratios with increasing galacto-centric radius in the broad lines
(Fig.~\ref{fig:NIIHa_radius}), with ([N{\sc ii}]\,/\,H$\alpha$)$_{\rm
  BL}$\,=\,($-$0.0253\,$\pm$\,0.004)\,$r_{\rm
  kpc}$\,+\,(0.47\,$\pm$\,0.02).  In contrast, the gradient in the
ratio of [N{\sc ii}]\,/\,H$\alpha$ emission line intensities with
galacto-centric radius is shallow, with ([N{\sc
    ii}]\,/\,H$\alpha$)$_{\rm
  NL}$\,=\,($-$0.006\,$\pm$\,0.003)\,$r_{\rm
  kpc}$\,+\,(0.22\,$\pm$\,0.02) (Fig.~\ref{fig:NIIHa_radius}).  The
gradient in [N{\sc ii}]\,/\,H$\alpha$ with radius for the narrow-line
emission is comparable to the negative, but shallow metallicity
gradients seen in individual high-redshift galaxies
\citep{Yuan11,Queyrel12,Stott14}.

Considering instead the [S{\sc ii}] emission line, in the same
  radial bins we do not detect any significant variation in the
  narrow-line ratio of S2\,=\,$I_{\rm 6716}$\,/\,$I_{\rm 6731}$, with
  an average of S2\,=\,1.45\,$\pm$\,0.15 (which is consistent with
  that derived from the composite).  The emission line ratio of
  $I_{\rm 6716}$\,/\,$I_{\rm 6731}$ for the broad component displays a
  weak trend with galacto-centric radius.  However since the trend is
  only 2$\sigma$, (with
  S2\,$\propto$\,($-$0.006\,$\pm$\,0.003)\,$r_{\rm kpc}$), in all
  following sections we adopt a fixed emission line ratio for the
  $I_{\rm 6716}$\,/\,$I_{\rm 6731}$ doublet, with S2\,=\,1.4 and
  S2\,=\,1.3 for the narrow- and broad- lines respectively.

Since the [O{\sc i}]\,$\lambda$6300\AA\, line is weak compared to
H$\alpha$ we investigate the [O{\sc i}]\,/\,H$\alpha$ ratio only in
the inner- versus outer- regions.  The signal-to-noise of the [O{\sc
    i}] precludes decomposing the [O{\sc i}] into a broad and narrow
component.  However, in integrated properties, we find a similar trend
as seen in the [N{\sc ii}] such that the central regions have a higher
emission line flux ratio, with [O{\sc
    i}]\,/\,H$\alpha$\,=\,(0.038\,$\pm$\,0.003) compared to [O{\sc
    i}]\,/\,H$\alpha <$\,0.0031 in the outer regions.  Thus, the
[N{\sc ii}]\,/\,H$\alpha$, and [O{\sc i}]\,/\,H$\alpha$ gradients
appear to reflect the increased ionisation of the outflow in the
central regions.

%
%  Figure 4
%

A prediction of the superwind model is that there should be a strong
pressure gradient (an approximately constant pressure profile within
the ISM around the energy-injecting source and an outer region where
the outflowing wind expands freely and the pressure falls off as
1/$r^{2}$).

We convert this into a pressure assuming a constant temperature
(10$^4$\,K) and a total density three times that of the electron
density determined from the [S{\sc ii}] ratio (to account for the fact
that the [S{\sc ii}] lines are formed in the partially ionised zone of
the nebulae).  This suggests a pressure P\,/$k_{\rm
  B}$\,=\,1--3\,$\times$\,10$^6$\,K\,cm$^{-3}$ (100--300\,$\times$
that of the mid-plane of the Milky-Way) which is approximately
constant over the radius 1--10\,kpc.  This observation suggests that
pressures much higher than in the Milky-Way are maintained over a large
volume (up to several kpc$^3$) in the centers of these galaxies.  If
the pressure does decrease with radius in the regime where the winds
expand freely, this must occur on scales $>$10\,kpc.

Since outflows should expand preferentially along the path of the
steepest pressure gradient, they are expected to preferentially expand
out of the plane of the galaxy \citep{Lehnert96}.  In local galaxies,
the winds tend to be weakly collimated
\citep[e.g.\ ][]{Dahlem97,Chen10,Bordoloi11}.  To search for the same
effect in our KMOS sample, and so test whether the winds are
collimated, or whether they are spherically symmetric (or wide angle),
we split the sample into five bins of inclination, each bin
containing the same number of galaxies, and measure the FWHM of the
broad and narrow components.  We identify broad H$\alpha$ in all
cases, with FWHM ranging from 208\,$\pm$\,4\,km\,/\,s$^{-1}$ for
galaxies at low inclination, to 245\,$\pm$\,6\,km\,s$^{-1}$ for more
highly inclined galaxies (Fig.~\ref{fig:stack_radius}).  Although the
trend is only weak, it is consistent with several other studies which
have also shown that outflows tend to be bi-polar with wide opening
angles \citep{Martin12,Bordoloi11}.

%
%  Figure 5
%
\begin{figure*}
  \centerline{\psfig{file=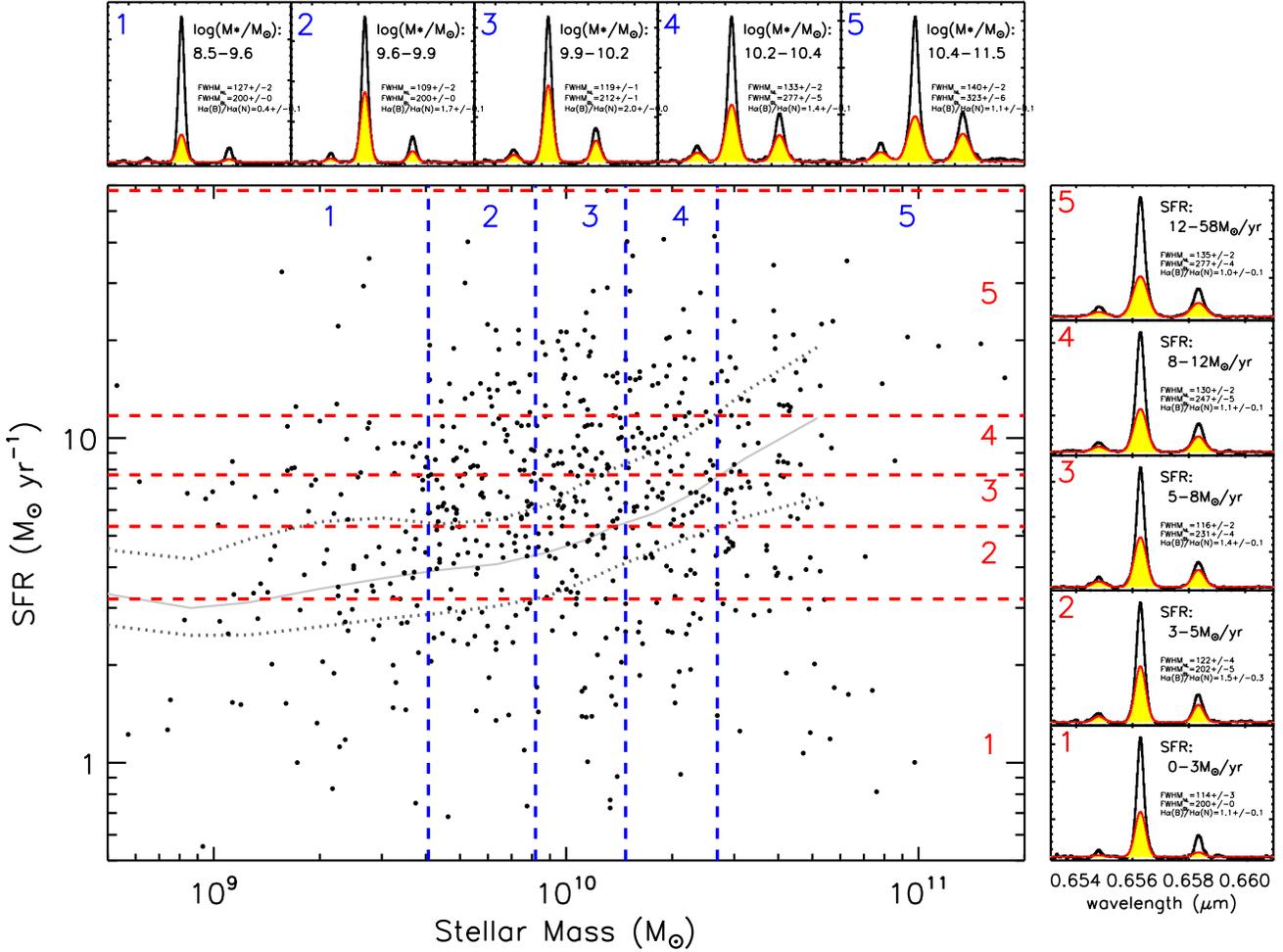,angle=90,width=7in}}
  \caption{Stacked spectra for the galaxies in our sample as a
    function of stellar mass and star formation rate.  The main panel
    shows the stellar masses and star formation rates for the sample.
    We bin the sample in to 5 bins of stellar mass and star formation
    rate (labeled 1--5 in each case).  Each bin contains the same
    number of galaxies, and are shown by the dashed lines.  The panels
    across the top and right-hand side of the figure show the stacked
    spectra for each bin, including the underlying broad line
    emission.  These plots demonstrate that the broad, underlying
    emission increases in FWHM with increasing stellar mass and
    star-formation rate.}
\label{fig:poster}
\end{figure*}

\section{Discussion}

With broad lines identified in the galaxy spectra, we can use the
broad line intensity and width to estimate the energetics of the
outflowing gas.

\subsection{Outflow Scaling Relations}
\label{sec:scalings}

Before estimating the mass outflow rates and kinetic energy in
  the wind from the composite stack, we first investigate the
  dependence of broad-line H$\alpha$ properties with galaxy
  properties.  We reiterate that there are significant systematic
  uncertainties in the absolute scaling of the mass outflow rates,
  kinetic energy and coupling efficiency we have derived.  However the
  relative scaling (e.g.\ with mass or star formation
  rate) should be much more reliable.  The most straight-forward prediction
of the superwind theory is that the properties of the emission-line
gas should be related to the star-formation rate of the galaxies since
the ``roots'' of the outflows are located in the OB associations and
associated supernovae events.  Other properties, such as rotation
speed and axial ratio are expected to correlate more weakly.  For
example, \citet{Hopkins12c} predict that the mass outflow rate from
starburst-driven superwinds scales linearly with star-formation rate.
Indeed, \citet{Martin12} used low-ionisation absorption lines to show
that the velocity of the out-flowing gas scales with star-formation
rate in galaxies at $z$\,=\,0.4--1.4, whilst \citet{Genzel14} used a
sample of high-redshift, massive, star-forming galaxies and AGNs to
suggests that the outflow velocity scales with galaxy mass and
star-formation rate.

\begin{figure*}
  \centerline{\psfig{file=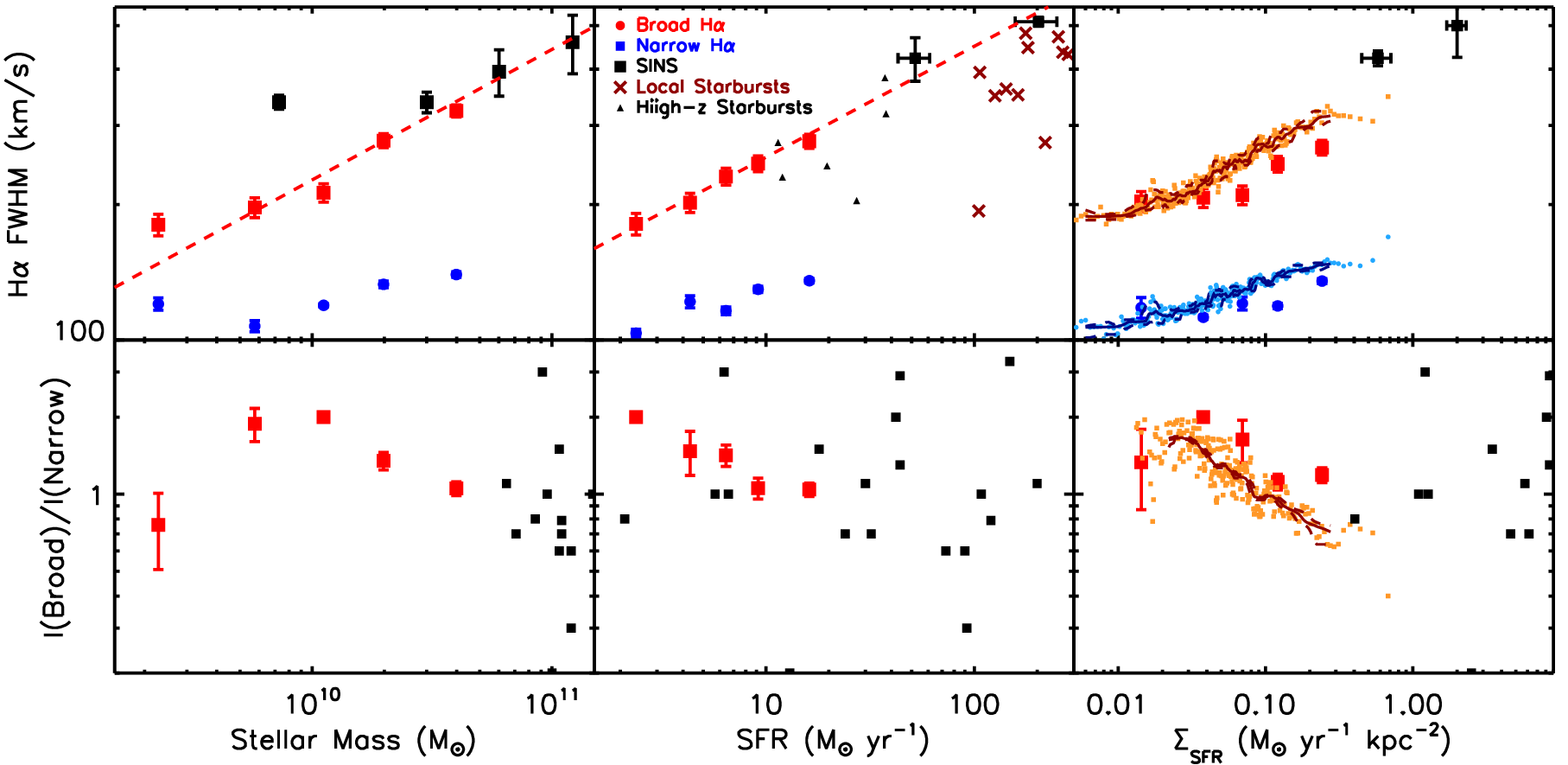,angle=90,width=6in}}
\caption{Physical properties of the H$\alpha$ broad lines as a
  function of mass, star formation rate and star formation surface
  density for the galaxies in our sample.  {\it Top Left:} Line width
  versus stellar mass for the galaxies in our sample compared to the
  SINS+KMOS$^{\rm 3D}$ survey of broad H$\alpha$ in $z$\,=\,1--2
  galaxies from \citet{Newman12} and \citet{Genzel14} and DEEP2
  galaxies from \citet{Martin12}.  The dashed line shows the outflow
  velocity as a function of mass with
  FWHM\,$\propto$\,M$_\star^{0.29}$.  {\it Top Center:} H$\alpha$ line
  width (FWHM) as a function of star formation rate.  The more
  actively star-forming galaxies tend to higher broad-line widths,
  increasing from FWHM\,$\sim$\,180\,km\,s$^{-1}$ to 350\,km\,s$^{-1}$
  between SFR\,=\,1--2\,M$_\odot$\,yr$^{-1}$ to
  $\gsim$\,30\,M$_\odot$\,yr$^{-1}$.  We also plot the variation in
  narrow-line H$\alpha$ FWHM.  Similar trends of increasing broad-line
  widths with star-formation rate have also been seen in H$\alpha$
  \citep[e.g.\ ][]{Newman12,Genzel14} and in low ionisation lines
  \citep{Martin12}.  The dashed line shows a power-law fit to the
  broad-line H$\alpha$ velocities in with
  FWHM\,$\propto$\,SFR$^{0.24}$.  {\it Top Right:} Broad line
  H$\alpha$ FWHM as a function of star formation surface density.  The
  large points denote the galaxy-averaged values, which show an
  increasing broad line FWHM with increasing star formation surface
  density.  We rank the pixels in each datacube by their individual
  star formation surface densities and plot the broad-line FWHM using
  bins that each contain 100 spectra (small points).  The red dashed
  solid line shows the running median through those data (with dashed
  lines denoting the central 68\% of the distribution).  In this plot,
  both the galaxy-averaged and pixel-to-pixel values show show a
  strong trend in broad line FWHM that increases strongly with
  increasing star formation surface density.  {\it Lower Left:} Broad
  H$\alpha$ emission line flux fraction ($f_{\rm broad}$\,/\,$f_{\rm
    narrow}$) as a function of stellar mass.  The broad H$\alpha$
  fraction is high, with $f_{\rm broad}$\,/\,$f_{\rm narrow}$\,=\,1--2
  across a decade of stellar masses.  {\it Lower Middle:} Broad
  H$\alpha$ fraction as a function of star formation rate, which shows
  a similarly strong broad\,/\,narrow emission line flux ratio.  {\it
    Lower Right:} Broad\,/\,Narrow H$\alpha$ emission line flux ratio
  as a function of star formation surface density.  We plot the galaxy
  average values (large solid points), and also the measurements from
  the ranked pixels from each datacube, with each measurement derived
  from 100 spectra.  The red dashed solid line shows the running
  median through those data (with dashed lines denoting the central
  68\% of the distribution).}
\label{fig:stack_obs_props}
\end{figure*}

To investigate how the outflow properties depend on the global galaxy
properties in our sample, we divide the sample into five bins of mass
and five bins of star-formation rate (with equal numbers of galaxies
in each bin) and show the resulting spectral stacks in
Fig.~\ref{fig:poster}.   Since sub-dividing the sample reduces the
  signal-to-noise, we fix the velocity offset between the narrow- and
  broad- components for all line to zero (although we note that a free
  fit returns velocity offsets $<$10\,km\,s$^{-1}$ in all cases).  We
  also fix the ratios of the broad- emission line intensities of the
  [S{\sc ii}] doublet to their average values from the stack of all
  galaxies (\S~\ref{sec:SpatialExtent}), but continue to allow
  variation in the relative contribution of the broad/narrow lines.
  In all of the following sections, the quoted errors on the line
  ratios and/or velocity widths incorporate the covariance between
  the fitted parameters.  As the spectra in Fig.~\ref{fig:poster}
show, there appears to be a trend of increasing broad-line H$\alpha$
FWHM and intensity with both mass and star formation rate, which is
also shown in Fig.~\ref{fig:stack_obs_props}.  In the local Universe,
the velocity of outflowing gas has also been shown to correlate with
star-formation rate, with a factor $\sim$\,30 change in velocity
observed over a range of four orders of magnitude of SFR ($v_{\rm
  wind}\propto$\,SFR$^{0.3}$; \citealt{Rupke05,Martin05}).  Indeed,
this figure shows that the high-redshift galaxies in our sample also
tend to have broader lines with increasing star-formation rate
(increasing from FWHM\,$\sim$\,180\,km\,s$^{-1}$ to 350\,km\,s$^{-1}$
between SFR\,=\,1--2\,M$_\odot$\,yr$^{-1}$ to
$\gsim$\,30\,M$_\odot$\,yr$^{-1}$) and mass (increasing from
FWHM\,$\sim$\,180\,km\,s$^{-1}$ to 323\,km\,s$^{-1}$ between
M$_\star\sim$\,(0.2--4)\,$\times$\,10$^{10}$\,M$_\odot$).  Although we
do not have the same dynamic range in star-formation rate as the local
samples, a power-law fit to the broad-line H$\alpha$ velocities in
Fig.~\ref{fig:stack_obs_props} yields
FWHM\,$\propto$\,SFR$^{0.24\pm0.08}$ and
FWHM\,$\propto$\,M$_\star^{0.29\pm0.08}$.

In Fig.~\ref{fig:stack_obs_props} we also plot the FWHM and
broad\,/\,narrow line intensity ratio of the underlying broad-line
H$\alpha$ with star formation surface density.  We plot the
galaxy-averaged values, but we also rank the pixels in each datacube
by their individual star formation surface density and then calculate
the broad-line FWHM using bins that each contain 100 spectra.  Both the
galaxy-averaged and pixel-to-pixel values show show a strong trend in
broad line FWHM that increases strongly with increasing star formation
surface density.

%
%  Figure 6
%

Similar trends of increasing broad-line widths with star-formation
rate have also been seen in the stacked H$\alpha$ spectra of
$z$\,=\,1--2 galaxies from \citet{Newman12} and \citet{Genzel14} as
well as in $z\sim$\,1 star-forming galaxies selected from the DEEP2
survey \citep{Martin12}.  We note that the velocities measured in the
rest-frame UV\,/\,optical spectra from \citet{Martin12} use low
ionisation lines (such as Fe{\sc ii} and Mg{\sc ii}), measuring the
maximum velocity, $v_{\rm max}$ as the velocity at which the
absorption line depth plus 1-$\sigma$ noise is consistent with the
continuum.  To compare this value to our FWHM, we construct a set of
model spectra with an absorption line doublet of variable FWHM and
intensity appropriate for the DEEP2 sample and add noise such that the
signal-to-noise per pixel is 5--7\,$\sigma$ (similar to the data in
\citealt{Martin12}).  For a reasonable range in FWHM and S\,/\,N, we
estimate FWHM\,=\,1.6\,$v_{\rm max}$ and use this conversion when
comparing results from \citet{Martin12} in
Fig.~\ref{fig:stack_obs_props}.

In Fig.~\ref{fig:stack_obs_props} we also plot the broad line FWHM
with stellar mass and compare our results with recent similar
observations of broad H$\alpha$ in $z$\,=\,1--2 galaxies and AGNs from
\citet{Newman12} and \citet{Genzel14} along with those from
\citealt{Martin14}.  This shows how the H$\alpha$ FWHM increases with
stellar mass (and star formation rate).  (see also
\citealt{Forster-Schreiber19,Freeman19}).  As well as the broad
H$\alpha$ line width, in Fig.~\ref{fig:stack_obs_props} we also show
the broad\,/\,narrow H$\alpha$ flux ratio ($f_{\rm broad}$\,/\,$f_{\rm
  narrow}$) as a function of star-formation rate, star formation
surface density and stellar mass.  We will combine the broad H$\alpha$
luminosities and line-widths in the next section to determine what the
trends of increasing FWHM with stellar mass and star formation rate
seen in Fig.~\ref{fig:stack_obs_props} imply for the outflow
energetics.

\subsection{Outflow Properties: estimates of mass, energy and momentum}
\label{sec:mor_all}

The ionised gas mass (i.e.\ the gas that is emitting the
H$\alpha$) can be estimated assuming an electron temperature
$T$\,=\,10$^4$\,K following \citealt{Osterbrock06} using the relation

\begin{equation}
  \frac{M_{\rm gas}}{2.38\times10^8\,{\rm M}_\odot}\,=\,\left(\frac{L_{H\alpha}}{10^{43}\,{\rm erg\,s^{-1}}}\right)\,\left(\frac{n_{\rm e}}{100\,{\rm cm^{-3}}}\right)^{-1}
\end{equation}

This relation has been used in a number of studies of galaxies
  involving H$\alpha$ (or H$\beta$) emission, although the
  normalisation of the co-efficient varies depending on the choice of
  assumptions \citep{Holt06,Genzel11,Liu13,RZ13,Harrison14}.  Since we
  do not have H$\beta$ measurements for this sample (to correct for
  the dust extinction in the emission lines), we adopt a simple
  approach to correct the observed broad H$\alpha$ luminosity to total
  H$\alpha$ luminosity.  For simplicity, we adopt the median
  reddenning for the KROSS sample (derived from the best-fit SED to
  the broad-band photometry; A$_{\rm v}$\,=\,1.10\,$\pm$\,0.07;
  \citealt{Harrison17}), and use the relation between stellar and
  gas-phase reddening from \citet{Wuyts13}; $A_{\rm gas}$\,=\,$A_{\rm
    v}$\,$\times$\,(1.9\,$-$\,0.15\,$A_{\rm v}$).  This calibration
  suggest an average gas-phase extinction at H$\alpha$ of $A_{\rm
    H\alpha}$\,=\,2.0\,$\pm$\,0.1 which is also consistent with the
  estimates from the far-infrared \citep[e.g.\ ][]{Thomson17}.  For
  the stack of all galaxies in our sample, we estimate an average
  ionised gas mass in the outflow of M$_{\rm
    gas}$\,=\,1--6\,$\times$\,10$^8$\,M$_\odot$ assuming $n_{\rm
    e}$\,=\,75\,cm$^{-3}$ (as derived from the [S{\sc ii}] doublet
  emission line ratio in \S~\ref{sec:extent}).  We note that this mass
  should be considered a lower limit since the ionised gas will only
  make up a fraction of the {\it total} out-flowing gas content.
  
%
%  Figure 7
%
\begin{figure*}
  \centerline{\psfig{file=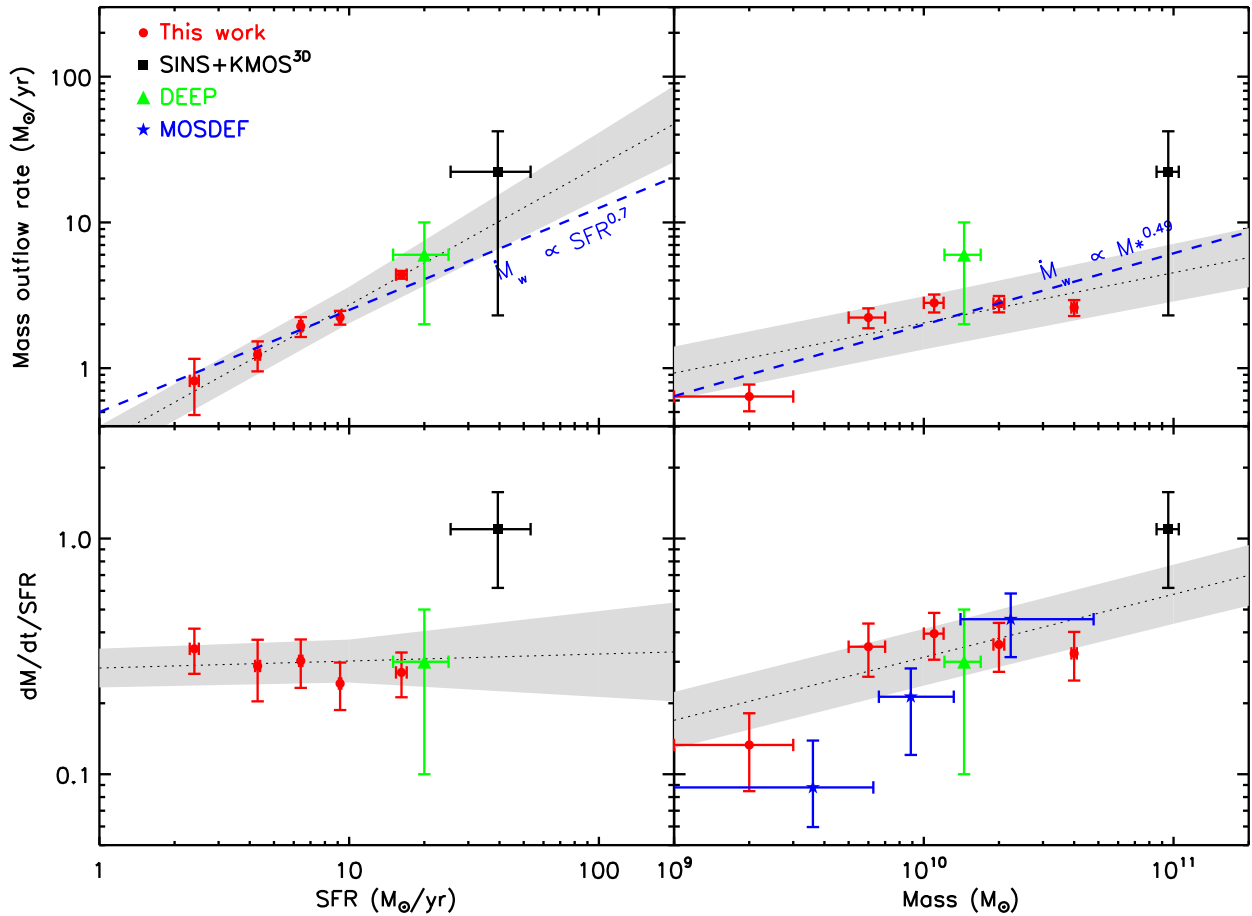,angle=90,width=6.5in}}
\caption{Inferred physical properties of the outflow as a function of
  galaxy properties.  {\it Top Left:} Mass outflow rate as a function
  of star formation rate for the galaxies in our sample.  The bins of
  star-formation rate are defined by equal numbers of galaxies per bin
  (see Table~1).  We also overlay the estimates of the mass outflow
  rates for high-redshift star-forming galaxies from \citet{Genzel14}
  and \citet{Freeman19} (both of which are also derived using the
  broad lines from composite spectra around the redshifted H$\alpha$).
  For consistency with our data, we have scaled the mass outflow rates
  from \citet{Freeman19} to the same spatial extent as those assumed
  in our observations.  We also include the measurements from
  \citet{Martin12} for high-redshift star-forming galaxies from the
  DEEP2 survey.  The data appear to show a trend of increasing mass
  outflow rate with star-formation rate with the form
  $d$M\,/\,$dt\propto$\,SFR$^{0.86\pm0.15}$ (dotted-line plus grey
  region), which is consistent with model predictions from
  \citet{Hopkins12c} which has the form $\dot{M}_{\rm
    wind}\propto$\,SFR$^{0.7}$. {\it Top Right:} Mass outflow rate as
  a function of stellar mass (with each stellar mass bin containing
  equal numbers of galaxies).  The trend of increasing mass outflow
  rate with stellar mass following $\dot{M}_{\rm wind}\propto
  M_\star^{0.49}$ from \citet{Hopkins12c} is in reasonable agreement
  with the data, which suggest
  $d$M\,/\,$dt\propto$\,SFR$^{0.34\pm0.14}$ (dotted line).  {\it
    Bottom Left:} Mass loading ($\dot{M}_{\rm wind}$\,/\,SFR) as a
  function of star formation rate.  The average mass loading is
  approximately constant over the star-formation rate range of our
  data, with $d$M\,/\,$dt$\,/\,SFR\,=\,0.3\,$\pm$\,0.1 and
  $d$M\,/\,$dt$\,/\,SFR\,$\propto$\,SFR$^{-0.07\pm0.14}$.  {\it Bottom
    Right:} The mass loading as a function of stellar mass.  In the
  fit, we include all of our data-points as well as the comparison
  samples in the fit, and identify a (weak) trend of increasing mass
  outflow rate with stellar mass.  The mass loading has a scaling with
  $d$M\,/\,$dt$\,/\,SFR\,$\propto$\,M$_\star^{0.26\pm0.07}$.  }
\label{fig:stack_phys_props}
\end{figure*}

After estimating the mass in the outflow, we estimate the kinetic
energy in the outflow as $E_{\rm kin}$\,=\,0.5\,M$_{\rm gas}$\,v$_{\rm
  gas}^2$.  Following \citet{Harrison14} and \citet{Liu13} we assume
that all of the ionised gas in the broad component is involved in the
outflow, with a bulk flow velocity of $v_{\rm gas}$\,=\,FWHM$_{\rm
  H\alpha}$\,/\,2 (suitable for a wide-angle bi-cone outflow or
spherically symmetric model).  For the composite stack in
Fig.~\ref{fig:stack_all}, this leads to a kinetic energy of
(1--8)\,$\times$\,10$^{56}$\,erg.  Assuming a maximum spatial extent
of 10\,kpc (Fig.~\ref{fig:stack_radius}) and a continuous outflow, the
outflow time is $t_{\rm wind}\sim $\,40\,Myr and consequently the
outflow kinetic energy rate is $\dot{E}_{\rm
  kin}$\,=\,(0.8--6)\,$\times$\,10$^{41}$\,erg\,s$^{-1}$.

The inferred mass outflow rate can be expressed in terms of the broad
H$\alpha$ luminosity and line width, electron density and spatial
extent as

\begin{multline}
  \frac{dM}{dt}\left(\frac{{\rm M}_\odot}{{\rm yr}^{-1}}\right)\,=\,0.024\left(\frac{L_{H\alpha}}{10^{43}\,{\rm erg\,s^{-1}}}\right)\,\left(\frac{n_{\rm e}}{100\,{\rm cm^{-3}}}\right)^{-1}\times\\  
  \left(\frac{{\rm FWHM}_{H\alpha}}{{\rm km\,s^{-1}}}\right)\,\left(\frac{10\,{\rm kpc}}{r}\right)
\end{multline}

\noindent For our sample, this suggests a mass outflow rate, dM$_{\rm
  wind}$\,/\,d$t$\,=\,3--18\,M$_\odot$\,yr$^{-1}$ (and so a mass
outflow rate per unit star formation rate of d$M_{\rm wind}$\,/\,d$t$\,/SFR less than unity).  An
alternative approach is to assume a spherical volume of out-flowing
ionised gas which gives a mass outflow rate of $\dot{M}_{\rm
  wind}$\,=\,3\,M$_{\rm gas}\,v_{\rm wind}\,/\,r$ and $\dot{E}_{\rm
  kin}$\,=\,0.5\,$\dot{M}$\,($v_{\rm wind}^{2}$\,+\,$\sigma^2$)
where $\sigma$ is the velocity dispersion and $v_{\rm wind}$ is the
outflow velocity.  To match local samples as closely as possible
\citep[e.g.\ ][]{RZ13}, we adopt $v_{\rm wind}$\,=\,FWHM\,/\,2, the
velocity dispersion to be $\sigma$\,=\,FWHM\,/\,2.355, the maximum
extent of the outflow to be a radius $<$\,10\,kpc and again use
$n_{\rm e}$\,=\,75\,cm$^{-3}$.  Using this approach, we obtain a mass
outflow rate of $\dot{M}_{\rm wind}$\,=\,1--8\,M$_\odot$\,yr$^{-1}$
and $\dot{E}_{\rm
  kin}$\,=\,(3--8)\,$\times$\,10$^{40}$\,erg\,s$^{-1}$.  This range in
outflow kinetic energy rate is in agreement with the range of values
using the first method.  Although we caution that there are
  potentially significant systematic uncertainties in our estimates
  \citep[e.g.\ see also][]{Harrison18}, these mass outflow rates and
  kinetic energy estimates are similar to those derived for local
  starbursts using absorption line studies for galaxies with comparable
  far-infrared luminosities to those studied here \citep{Heckman00}.
  However, we reiterate that both methods we employ are likely to
  provide lower limits on the total mass of the outflow as we are only
  observing the line-emitting (ionised) gas.  Most of the outflow is
  likely to be in the atomic or molecular phases that are cooler than
  the ionised material \citep{Rupke05,Rupke13,Walter02}.  Hence the
  total mass involved in the outflow could be much higher (see also
  e.g.\ \citealt{Greene11}).

We examine how the radiation pressure and supernovae energy compare to
the energy in the outflows.  First, we assume that the energy from
supernovae per solar mass of stars formed per year is
1\,$\times$\,10$^{49}$\,erg\,yr$^{-1}$ (applicable for stellar ages
$\gsim$\,40\,Myr and following
\citealt{Leitherer99,Veilleux05})\footnotemark.  Thus, for our
composite sample with a median star-formation rate of
SFR\,=\,7\,$\pm$\,1\,M$_\odot$\,yr$^{-1}$, the total energy available
from SNe is $\sim$\,7\,$\times$\,10$^{49}$\,erg, and the coupling
efficiency of the SNe to the wind is $\epsilon_{\rm
  SNe}$\,=\,0.7--3\%.  \footnotetext{If we instead used the supernovae
  energy from \cite{DallaVecchia08}, these values would be a factor
  $\sim $\,2\,$\times$ lower.}  We also compare the momentum rate in
the outflow ($\dot{P}$) to the radiation momentum rate ($L$\,/\,$c$).
The average infrared luminosity for the full sample (measured from the
stacked \emph{Herschel}\,/\,PACS+SPIRE photometry;
e.g.\ \citealt{Stott16}) is $L_{\rm
  IR}\sim$\,1\,$\times$10$^{11}$\,L$_\odot$, which suggests a
radiative momentum flux rate of $L_{\rm
  IR}$\,/\,c\,=\,(6\,$\pm$\,2)\,$\times$\,10$^{35}$\,erg\,m$^{-1}$.
In comparison to the momentum rate of the outlow (with
$\dot{P}$\,=\,$\dot{M}_{\rm wind}$), the momentum ratio is in the
range $\dot{P}$\,/\,$L_{\rm IR}$\,=\,7--600 (with a median of
$\dot{P}$\,/\,$L_{\rm IR}$\,=\,27), and thus not indicative of
momentum driven outflows \citep{Dekel13}.

\subsection{Outflow Properties: dependence on mass and star formation rate}
\label{sec:mor_mass_sfr}

Whilst it is useful to estimate the energetics of the outflows in the
full sample of 529 galaxies, there are significant systematic
  uncertainties in the absolute values derived.  However, numerical
models suggest that the outflow energetics scale (non linearly) with
circular velocity, star formation and gas density
\citep[e.g.][]{Hopkins12c,Barai15}, and so a useful test can be
  performed by comparing these quantities between sub-samples as a
  function of stellar mass and star-formation rate.  An in \S2, we therefore
sub-divide the stacks in to smaller subsamples to investigate how the
energetics of the wind (kinetic energy, mass loading) depend on galaxy
properties to test the predictions of the superwind theory -- that the
energetics of the wind should correlate with the star-formation rate
of the galaxy.

In Fig.~\ref{fig:stack_phys_props} we show how the physical properties
of the outflows correlate as a function of galaxy properties derived
using the line widths and luminosities from
Fig.~\ref{fig:stack_obs_props} (see also Table~2).  First we derive
the mass outflow rate as a function of star-formation rate and overlay
those calculated for high-redshift star-forming galaxies by
\citet{Genzel14} (which are also derived using the broad lines from
composite spectra around the redshifted H$\alpha$) and those for
DEEP\,2 galaxies from \citet{Martin12}.  To be consistent with our
sample, from the SINS+KMOS$^{\rm 3D}$ comparison sample from Genzel et
al.\ we exclude (candidate) AGNs and/or galaxies with [N{\sc
    ii}]\,/\,H$\alpha>$\,0.7 and adopt a maximum spatial extent for
the outflow of 2\,$\times$\,HWHM.  Fig.~\ref{fig:stack_phys_props}
shows an increasing trend of mass outflow rate, $d$M\,/\,d$t$ with
star-formation rate.  The scaling between star-formation rate and mass
outflow rate suggested by the data,
$d$M\,/\,$dt\propto$\,SFR$^{0.86\pm0.15}$ is consistent with
predictions from hydrodynamical models, which predict a sub-linear
power-law relation ($\dot{M}_{\rm wind}\propto\dot{M_\star}^{0.7}$)
that implies a lower efficiency at higher star-formation rates
\citep{Hopkins12c}.  We also plot the relation between mass outflow
rate and stellar mass in Fig.~\ref{fig:stack_phys_props} which also
shows a sub-linear trend increasing with increasing mass.  The scaling
between stellar mass and mass outflow rate is
$d$M\,/\,$dt\propto$\,SFR$^{0.34\pm0.14}$, which is also similar to
the predictions from \citet{Hopkins12c}, $\dot{M}_{\rm
  wind}\propto\dot{M_\star}^{0.49}$.

The dependence of the mass loading ($d$M\,/\,$dt$\,/\,SFR) on stellar
mass or star formation is less significant.
Fig.~\ref{fig:stack_phys_props} shows that the mass loading is
approximately constant over star-formation rate range of our data,
with $d$M\,/\,$dt$\,/\,SFR\,=\,0.3\,$\pm$\,0.1, although there appears
to be a weak dependence on stellar mass, with
$d$M\,/\,$dt$\,/\,SFR\,$\propto$\,M$_\star^{0.26\pm0.07}$.  In the
latter fit, we also include similar constraints for $z\sim$\,1 star
forming galaxies from the recent MOSDEF survey from \citet{Freeman19}
(who also measure the underlying H$\alpha$ broad emission line
properties).  We note however that for consistency, we have scaled
their derived mass outflow rates to assume the same spatial extent as
our data.

%
% Table 2
%
\begin{table*}
{\footnotesize
\begin{center}
\caption{Derived Properties of the Outflows as a function of star formation rate and stellar mass}
\begin{tabular}{lcccccc}
\hline
\hline
Stack            & $M_{\rm wind}$        & d$M_{\rm wind}$\,/\,d$t$ & d$M_{\rm wind}$\,/\,d$t$\,/\,SFR
                                                                                  & E$_{\rm k,wind}$        & E(SNe)\,/\,E$_{\rm k,wind}$ & ($L$\,/\,$c$)\,/\,$\dot{P}$ \\
                 & (10$^{8}$\,M$_\odot$) & (M$_\odot$\,yr$^{-1}$)  &                  &    (10$^{54}$\,erg)   &         (\%)        &    (\%)                     \\
\hline
All              & 0.89\,$\pm$\,0.06  &  2.53\,$\pm$\,0.54  &  0.38\,$\pm$\,0.08  &  70\,$\pm$\,2  &  2.9\,$\pm$\,0.6   &  0.4\,$\pm$\,0.1  \\
\hline
SFR-1            & 0.44\,$\pm$\,0.03  &  0.82\,$\pm$\,0.17  &  0.34\,$\pm$\,0.07  &  15\,$\pm$\,1   &  1.1\,$\pm$\,0.2  &  0.6\,$\pm$\,0.1  \\
SFR-2            & 0.60\,$\pm$\,0.13  &  1.24\,$\pm$\,0.36  &  0.29\,$\pm$\,0.08  &  25\,$\pm$\,1   &  1.2\,$\pm$\,0.3  &  0.7\,$\pm$\,0.2  \\
SFR-3            & 0.83\,$\pm$\,0.10  &  1.94\,$\pm$\,0.45  &  0.30\,$\pm$\,0.07  &  44\,$\pm$\,2   &  1.6\,$\pm$\,0.4  &  0.5\,$\pm$\,0.1  \\
SFR-4            & 0.89\,$\pm$\,0.10  &  2.23\,$\pm$\,0.51  &  0.24\,$\pm$\,0.06  &  54\,$\pm$\,2   &  1.5\,$\pm$\,0.4  &  0.6\,$\pm$\,0.1  \\
SFR-5            & 1.55\,$\pm$\,0.10  &  4.38\,$\pm$\,0.92  &  0.27\,$\pm$\,0.06  &  120\,$\pm$\,15 &  2.1\,$\pm$\,0.5  &  0.5\,$\pm$\,0.1  \\
\hline
Mass-1           & 0.35\,$\pm$\,0.10  &  0.64\,$\pm$\,0.22  &  0.13\,$\pm$\,0.05  &  11\,$\pm$\,1   &  0.4\,$\pm$\,0.2  &  1.6\,$\pm$\,0.6  \\
Mass-2           & 1.11\,$\pm$\,0.16  &  2.22\,$\pm$\,0.55  &  0.35\,$\pm$\,0.09  &  43\,$\pm$\,1   &  1.4\,$\pm$\,0.3  &  0.6\,$\pm$\,0.1  \\
Mass-3           & 1.30\,$\pm$\,0.03  &  2.80\,$\pm$\,0.57  &  0.39\,$\pm$\,0.09  &  59\,$\pm$\,2   &  1.8\,$\pm$\,0.4  &  0.5\,$\pm$\,0.1  \\
Mass-4           & 0.98\,$\pm$\,0.06  &  2.77\,$\pm$\,0.58  &  0.36\,$\pm$\,0.08  &  76\,$\pm$\,3   &  2.7\,$\pm$\,0.6  &  0.4\,$\pm$\,0.1  \\
Mass-5           & 0.79\,$\pm$\,0.06  &  2.61\,$\pm$\,0.56  &  0.33\,$\pm$\,0.08  &  83\,$\pm$\,3   &  3.4\,$\pm$\,0.8  &  0.5\,$\pm$\,0.1  \\

\hline
\label{table:stack}
\end{tabular}
\end{center}
\noindent{\footnotesize Notes: Estimates of the mass outflow rates and
  their energetics as a function of stellar mass and star formation
  rate.  These are calculated using the extinction corrected, broad
  line H$\alpha$ luminosity and FWHM as discuss in section 3.2.
  Error-bars reflect uncertainties on the line fits.  }}
\end{table*}

Finally, we attempt to infer the fate of the outflowing gas by
investigating whether the outflows have sufficient speed to escape the
gravitational potential of the galaxies, or whether the gas is likely
to return to the galactic disk as a ``fountain'' flow.  For an
isothermal gravitational potential that extends to a maximum radius
$r_{\rm max}$ and has a circular velocity $v_{\rm c}$, the escape
velocity at a radius $r$ is given by:

\begin{equation}
  v_{\rm esc}\,=\,(2\,v_{\rm c}^2\,[1\,+\,{\rm ln}(r_{\rm max}\,/\,r)]\,)^{0.5}
\end{equation}

\begin{figure}
  \centerline{\psfig{file=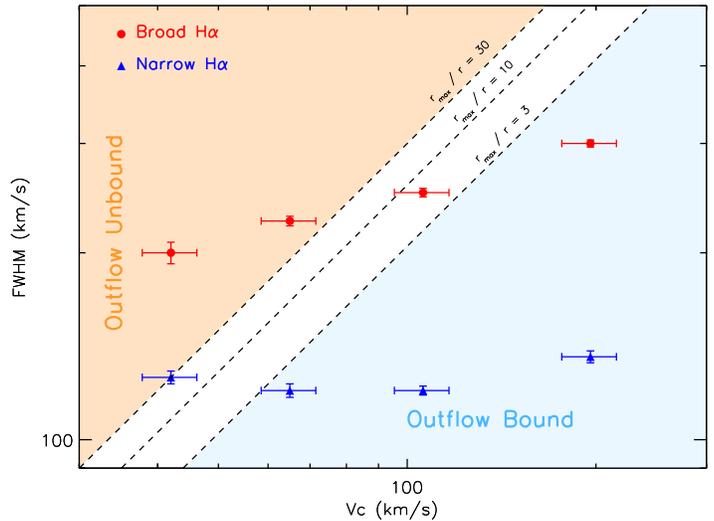,angle=90,width=4in}}
\caption{FWHM of the broad- and narrow- line H$\alpha$ as a function of
  (inclination corrected) circular velocity.  The dashed lines denote
  the gas velocity required to escape the gravitational potential
  assuming an isothermal gravitation potential that extends to a
  maximum radius of $r_{\rm max}$ and has a virial velocity $v_{\rm c}$
  at radius $r$.  The escape velocity is given by $v_{\rm
    esc}$\,=\,(2\,$v_{\rm c}^2$(1\,+\,$ln(r_{\rm max}/r)$))$^{0.5}$ and
  in the plot we show three values of $r_{\rm max}\,/\,r$\,=\,3, 10 and
  30.  Above $r_{\rm max}$\,/\,$r$\,=\,30, the outflowing gas
  is likely to escape the gravitational potential of the galaxy, and
  the lowest mass galaxies in our sample (identified by the lowest
  circular velocity, $V_{\rm c}$) meet this criteria.  In contrast, in
  the highest mass (highest $V_{\rm c}$) galaxies, the outflow does not
  appear to have sufficient velocity to escape the potential and will
  be retained.  }
\label{fig:stack_Vc}
\end{figure}

Thus, $v_{\rm esc}$\,=\,2.5\,$v_{\rm c}$ for $r_{\rm max}$\,/\,$r\sim
$\,9 (i.e. $r_{\rm max}$\,=\,100\,kpc and $r$\,=\,10\,kpc).  In
Fig.~\ref{fig:stack_Vc} we show the broad (and narrow) line H$\alpha$
FWHM as a function of (inclination corrected) circular velocity for
the galaxies in our sample.  This figure shows the FWHM of the broad
H$\alpha$ increases with circular velocity, and we derive a scaling
FWHM\,$\propto v_{\rm c}^{0.21\,\pm\,0.05}$.  This scaling is
significantly sub-linear and, as Fig.~\ref{fig:stack_Vc} shows, the
galaxies with lowest $v_{\rm c}$ (shallower potentials) have gas with
FWHM greater than our estimates of the velocity required to escape the
galaxy.  Conversely those at higher $v_{\rm c}$ (and hence higher mass
and deeper potentials) have gas with FWHM that is below that required
to escape the galaxy.  Of course, here we are comparing the FWHM of
the broad line with the escape velocity required for a single cloud of
gas, and therefore some care must be taken when comparing these
quantities outright.  Nevertheless, a significant fraction of the gas
in the lowest mass galaxies should have sufficient velocity to escape
the gravitational potential, whilst in the highest mass galaxies, most
of the gas will be retained, most likely flowing back on to the galaxy
disk (as also can also be seen in Fig.~\ref{fig:stack_obs_props}).

%
%  Figure 8
%

\section{Conclusions}
\label{sec:conc}

By exploiting KMOS observations of 529 star-forming galaxies at
$z\sim$\,1, we have investigated the average properties of starburst
driven, galaxy-scale outflows.  Our sample spans a range of mass and
star-formation rate, from
M$_\star$\,=\,0.5--5\,$\times$\,10$^{10}$\,M$_\odot$ and
SFR\,=\,1.5--28\,M$_\odot$\,yr$^{-1}$.  We measure the two dimensional
velocity field for each galaxy, model and subtract the galaxy
dynamics, and then stack the rest-frame optical spectra to search for
and measure the properties of the underlying, broad H$\alpha$
emission.  We identify broad H$\alpha$ emission along with forbidden
lines of [N{\sc ii}] and [S{\sc ii}].  The presence of broad forbidden
lines suggests that the outflows are not confined to a broad line
region around an AGN (and indeed, we remove galaxies in our sample
that display AGN characteristics in their optical spectra,
mid-infrared colours or X-ray emission).  We show that the broad
emission is spatially extended across at least 10\,kpc in projection.

The composite spectra has a reddening corrected star-formation rate
SFR\,=\,7\,M$_\odot$\,yr$^{-1}$ and
$M_\star$\,=\,(0.9\,$\pm$\,0.1)\,$\times$\,10$^{10}$\,M$_\odot$, and
from the broad H$\alpha$ we derive mass outflow rates of
2--10\,M$_{\odot}$\,yr$^{-1}$.  By comparing the kinetic energy in the
wind with the energy released by supernovae, we estimate a coupling
efficiency of $\lsim $3\%. This coupling efficiency is also comparable
to that between the radiation pressure from star formation, suggesting
that both the radiation pressure and SNe have sufficient energy or
momentum to drive the outflow (if they are able to couple $\sim $\,3\%
of their energy to the gas).   Although there are systematic
  uncertainties in these mass outflow rates depending on the
  assumptions made, these coupling efficiencies are much less than
  unity and suggest that either the supernovae or radiation pressure
  should be able to drive the outflow.

We also investigate the dependence of wind energetics with global
galaxy properties (mass and star formation).  Although the data are
limited by the sample size per bin (and systematics in deriving
physical parameters), we show that the mass outflow rates increases
with star-formation rate and mass with
$d$M\,/\,$dt\propto$\,SFR$^{0.86\pm0.15}$ and
$d$M\,/\,$dt\propto$\,SFR$^{0.34\pm0.14}$.  Both of these scalings are
consistent with predictions from hydrodynamical models which predict
sub-linear power-law relations of
$d$M\,/\,$dt\propto\dot{M_\star}^{0.49}$ \citep{Hopkins12c}.  However,
the mass loading of the winds ($\dot{M}_{\rm wind}$\,/\,SFR) shows
little dependence on star formation rate (over the range covered by
our data), although we identify a weak trend with stellar mass such
that $d$M\,/\,$dt$\,/\,SFR\,$\propto$\,M$_\star^{0.26\pm0.07}$.

We investigate whether the outflows have sufficient speed to escape
the gravitational potential of the galaxies, or whether the gas is
more likely return to the galactic disk as a fountain flow.  The line
width of the broad H$\alpha$ increases with disk circular velocity
with a sub-linear scaling relation FWHM\,$\propto v_{\rm
  c}^{0.21\,\pm\,0.05}$.  In the lowest mass galaxies,
M$\lsim$\,10$^{10}$\,M$_\odot$ a significant fraction of the gas
should have sufficient velocity to escape the gravitational potential,
whilst in the highest mass galaxies, most of the gas will be retained,
most likely flowing back on to the galaxy disk at later times.

These results are based on average values for (sub)-populations binned
by mass, star-formation rate, or circular velocity.  Of course, there
are likely to be variations in individual objects in terms of (e.g.)
sizes, mass, star formation rate or electron densities
\citep[e.g.][]{Rose18}.  To make the same measurements in individual
galaxies will require extremely deep exposures ($\gsim$\,100\,hr) per
galaxy, which the next generation of KMOS surveys should
provide.

\section*{acknowledgments}

We would like to thank the referee for their constructive report which
significantly improved the content and clarity of the paper.  We
gratefully acknowledge STFC through grant ST/L00075X/1.  IRS and AMS
acknowledge the Leverhume foundation.  IRS and AT also acknowledges
the ERC Advanced Grant programme {\sc dustygal}.  IRS acknowledges a
Royal Society Wolfson Merit Award.  The KMOS and \emph{Herschel} data
used in this paper are available through the ESO and HEDAM archives.
The KMOS data in this paper were taken as part of programs 60.A-9460,
092.B-0538, 093.B-0106, 094.B-0061, and 095.B-0035.

%\bibliographystyle{apj}
%\bibliography{/Users/ams/Projects/ref}

\end{document}